\documentclass[12pt, a4paper]{article}
\usepackage{float}
\usepackage{multirow}
\usepackage{amssymb,amsmath, amsfonts}
\usepackage{lscape}
\usepackage{enumerate}
\usepackage{amsthm}
\usepackage[T1]{fontenc}
\usepackage{hyperref}
\usepackage[utf8]{inputenc}
\usepackage{lmodern}
\usepackage[english]{babel}
\usepackage{csquotes}
\usepackage{mathtools}

\usepackage{xcolor}
\numberwithin{equation}{section}
\usepackage{graphicx}
\usepackage{caption}
\usepackage{subcaption}
\usepackage{upgreek}
\newtheorem{theorem}{Theorem}[section]
\newtheorem{example}{Example}[section]

\newtheorem{remark}{Remark}[section]

\newtheorem{definition}{Definition}[section]

\begin{document}

\begin{center}
{\Large\bf Multivariate Information Measures: A Copula-based Approach} \\
\vspace{0.3in}

{\bf Mohd. Arshad$^{a}$, \bf Swaroop Georgy Zachariah$^{a}$\footnote{Corresponding author. E-mail
addresses: ~arshad@iiti.ac.in (M. Arshad), ~ashokiitb09@gmail.com (Ashok Kumar Pathak), ~swaroopgeorgy@gmail.com
(Swaroop Georgy Zachariah).} and \bf Ashok Kumar Pathak$^{b}$}
% \\
%$^a$ Department of Mathematics, Indian Institute of
%Technology Indore, India \\
% $^b$ Department of Mathematics and Statistics,
%Central University of Punjab, Bathinda,
%India
\end{center}

\noindent------------------------------------------------------------------------------------------------------------------------------
\begin{abstract}
Multivariate datasets are common in various real-world applications. Recently, copulas have received significant attention for modeling dependencies among random variables. A copula-based information measure is required to quantify the uncertainty inherent in these dependencies. This paper introduces a multivariate variant of the cumulative copula entropy and explores its various properties, including bounds, stochastic orders, and convergence-related results. Additionally, we define a cumulative copula information generating function and derive it for several well-known families of multivariate copulas. A fractional generalization of the multivariate cumulative copula entropy is also introduced and examined. We present a non-parametric estimator of the cumulative copula entropy using empirical beta copula. Furthermore, we propose a new distance measure between two copulas based on the Kullback-Leibler divergence and discuss a goodness-of-fit test based on this measure.
\end{abstract}
\vskip 2mm


\noindent------------------------------------------------------------------------------------------------------------------------------

%%%%%**************************** Section 1 ********************************************************
\section{Introduction}
Entropy is a well-known concept in information theory and has applications in different disciplines of the applied sciences, ranging from statistical mechanics to machine learning, finance, insurance, physics, chemistry, and reliability (see \cite{a1,a2,a3,a4,a5}). The construction and generalization of the different entropy variants have drawn a lot of interest recently from both theoretical and practical perspectives (see, \cite{hosseini2021discussion,f5,f6}). The early development of entropy credits to Shannon \cite{shannon1948mathematical} in order to quantify uncertainty in systems. For a discrete random variable $X$ with probability mass function $p_i=P(X_i=x_i), i=1,2,\dots,n$, the Shannon entropy is defined as
$$\text{S}(X)=-\sum_{i=1}^{n}p_i\ln{p_i}.$$

The differential entropy (DE) is a continuous analogue of the Shannon entropy. Let $X$ be an absolutely continuous random variable with density $f$. Then DE takes the following form
$$\text{DE}(X)=-\int_Xf(x)\ln{f(x)}dx.$$
The DE faces some limitations. For instance, it can take negative values for some random variables. Furthermore, its approximation using empirical distribution functions is quite challenging. To address these limitations, \cite{rao2004cumulative} introduced a new measure of uncertainty based on survival function, which is called as cumulative residual entropy (CRE). For a non-negative random variable \(X\) with survival function \(\Bar{F}(x)\), the CRE is defined as
\[
\text{CRE}(X) = -\int_0^\infty \Bar{F}(x) \ln \Bar{F}(x) \, dx.
\]

 One can observe that CRE can be considered a natural generalization of the differential entropy and is valid for both discrete and continuous types of random variables. The interesting feature of the CRE is that it can be estimated from the sample, and various asymptotic results are easy to establish. On a similar line,
 \cite{di2009cumulative} defined the cumulative entropy by considering cumulative distribution function (CDF) instead of survival function. 
Apart from these measure, there are various measure of the entropy has been proposed and studied in the recent literature. Some important among these includes \cite{renyi1961measures, tsallis1988possible, di2007weighted,mathai2007pathway}. 
\par Recent research has focused on developing information generating functions, which can generate a number of useful uncertainty and divergence measures. Golomb \cite{golomb1966information} defined an information generating function by $$SG_X(s)=\sum_{i=1}^{n}\left(p_i\right)^s, \ s\geq 1.$$
It may be observed that $SG_X(1)=1$ and the first derivative of $SG_X(s)$ at $s=1$ corresponds to negative of the Shannon entropy. Guiasu and Reischer \cite{guiasu1985relative} discussed the generating function for the relative entropy and showed that its first derivative at $s=1$ gives negative of the Kullback and Leibler distance for two probability distributions proposed by \cite{kullback1951information}. Fisher information generating function and associated results are reported in \cite{papaioannou2007some}. The cumulative information generating function is introduced by \cite{smitha2019dynamic}.
The generating function and non-parametric estimator for the CRE is reported in a recent work of \cite{smitha2023dynamic}. Recently, Saha and Kayal \cite{saha2024general} defined the general weighted information and relative information generating functions and discuss its mathematical properties. 
\par Apart from these works, considering the idea of fractional calculus, fractional variants of the various information measures have been proposed, which extends a number of entropy existing in the literature. Various characteristics of fractional calculus allow the measure to handle long-range phenomena and non-local dependence in certain complex random systems. Ubraico \cite{ubriaco2009entropies} generalized the Shannon entropy using fractional calculus. The fractional version of the Shannon entropy is given by
\begin{equation}\label{fse}
    S_{r}(X)=\sum_{i=1}^{n}p_i\left(-\ln{p_i}\right)^r, \  0\leq r\leq 1.
\end{equation}
For $r=1$, $S_{r}(X)$ reduces to Shannon entropy. Xiong et al. \cite{xiong2019fractional} extended the idea for generalizing cumulative residual entropy and illustrated the application of fractional entropy for measuring the uncertainty in the financial data set. The authors also showed that fractional version reveal much information over classical one.
The fractional version of cumulative entropy were discussed in \cite{kayid2022some}.
For some more related work in this direction, one can refer to \cite{f1,f2,f3,f4,f5,f6}. 
\par In many real-life applications, multivariate data is often encountered. To quantify the associated uncertainty with multivariate random vectors, multivariate extensions of information measures are required. However, the literature addresses the uncertainty involved in the joint behavior of random vectors only to a limited extent. Some relevant works are discussed below.

Nadarajah and Zografos \cite{nadarajah2005expressions} discussed the bivariate Shannon and Rényi entropies. Their work was extended by Ebrahimi and Kirmani \cite{ebrahimi2007multivariate}, who obtained information measures for the residual lifetime of a bivariate random vector. Rajesh and Sunoj \cite{rajesh2009bivariate} introduced a vector-based bivariate residual entropy to measure the uncertainty involved in the remaining life of a bivariate random vector. Rajesh et al. \cite{rajesh2014bivariate} further extended the bivariate version of dynamic cumulative residual entropy proposed by Asadi and Zohrevand \cite{asadi2007dynamic}. Additionally, Kundu and Kundu \cite{c2017bivariate} extended the cumulative entropy proposed by Di Crescenzo and Longobardi \cite{di2007weighted} for a bivariate random vector and discussed its dynamic version.

\par In recent years, copula has emerged as a powerful tool for measuring the dependence among random variables. It connects the marginals to the joint distributions and allow us to develop a large class of multivariate distributions with desired dependence. Let $\mathbf{X}=\left(X_1,X_2,\dots,X_k\right)$ be a $k$-dimensional random variable having joint CDF $\mathbf{F}\left(x_1,x_2,\dots,x_k\right)$. Assume that each $X_i$ has CDF $F_i(x_i)$, $i=1,2\dots,k$. Sklar \cite{sklar1959fonctions} proved that $\mathbf{F}\left(\cdot\right)$ can be represented as a function of the marginal CDFs. i.e., there exist a function, called copula, $C:\mathbb{I}^k\rightarrow\mathbb{I}$ such that
\begin{equation*}
\mathbf{F}\left(\mathbf{x}\right)=C\left(F_1(x_1),F_2(x_2),\dots,F_k(x_k)\right),
\end{equation*}
where $\mathbf{x}=(x_1,x_2,\dots,x_k)\in\mathbb{R}^k$ and $\mathbb{I}=[0,1]$. If $F_i(\cdot);i=1,2,\dots,k$ are continuous then the underlying copula $C$ is uniquely determined. 
For more details see \cite{nelsen2007introduction}, \cite{durante2016principles}, \cite{chesneau2022note} and \cite{zachariah2024new}. Let $C$ be a $k$-dimensional absolutely continuous copula, then the copula density $D$ is defined as
$$D(\mathbf{u})=\dfrac{\partial^k C(u_1,u_2,\dots,u_k)}{\partial u_1 \partial u_2\dots \partial u_k},$$
where $\mathbf{u}=(u_1,u_2,\dots,u_k)$. It is to be noted that for every $k$-dimensional random vector $\bf{X}$ with joint probability density (PDF), $f(\bf{x})$ can be expressed as
\begin{equation}\label{pdf}
   f(\mathbf{x})= D\left(F_1(x_1),F_2(x_2),\dots,F_k(x_k)\right)\displaystyle\prod_{i=1}^{k} f_i(x_i),
\end{equation}
 where $f_i(\cdot)$ and $F_i(\cdot)$ is the marginal PDF and CDF of $X_i$, $i=1,2,\dots,k$, respectively. Ma and Sun \cite{ma2011mutual} proposed copula entropy defined as
\begin{equation*}
  \Delta\left(D\right)=-\int_{\mathbb{I}^k} D(\mathbf{u})\ln(D(\mathbf{u}))d\mathbf{u},
\end{equation*}
and showed that the copula entropy is negative of the mutual
information of a multivariate random vector. Since the copula density may not always exist, it is not a suitable
measure. Moreover, it is always non-positive. To overcome this limitation, we propose a multivariate cumulative copula entropy (CCE), which extend the bivariate CCE defined by \cite{sunoj2023survival}.
In many practical applications, multivariate data is often encountered, making it crucial to define multivariate CCE to quantify the uncertainty involving the dependence structure among random variables. Similar to CRE and copula entropy, the proposed CCE can be highly useful in various real-world scenarios such as image processing, financial engineering and hydrology (see \cite{rao2004cumulative}, \cite{xiong2019fractional}, \cite{ce1}, \cite{ce2}, \cite{ce3}). This paper discusses two specific applications of the multivariate CCE: the goodness-of-fit test for copula and the copula selection problem.  Considering the significance of information-generating functions and fractional order entropies in various practical contexts, it is relevant to study copula based entropies. This paper aims to study the copula based multivariate information measures and its associated properties, offering insights into their practical relevance. The main contribution of this paper are highlighted as follows.
 
\begin{enumerate}[1.]
    \item We propose the multivariate version of the CCE and discuss its mathematical properties, including bounds, stochastic orders, and convergence-related results. It is shown that CCE of the weighted arithmetic mean of copulas never exceeds weighted arithmetic mean of the CCE of copulas. 
    \item We propose the cumulative copula information generating function (CCIGF). It is shown that the first derivative of the CCIGF at $s = 1$ provides the negative of the CCE. Moreover, the proposed CCIGF is linearly related to the cumulative copula extropy proposed by \cite{saha2023copula}. 
    \item Fractional order entropies outperform classical entropies in many real-life applications (see \cite{xiong2019fractional} and \cite{kayid2022some}). Considering the importance of fractional order entropies, we also propose a fractional order of CCE.
    \item We propose a non-parametric estimator for the proposed CCE using the empirical beta copula and discuss its convergence.
    \item The Kullback-Leibler (KL) divergence measure is widely used in data analysis, including model selection criteria. We introduce a KL-based cumulative copula divergence, which is useful for copula selection problems. Additionally, we propose a goodness-of-fit test based on this new measure.
\end{enumerate}
The present paper is organized as follows. In Section \ref{sec2}, we discuss some mathematical properties of multivariate CCE and provide some examples. In Section \ref{sec3}, we introduce CCIGF and discuss some important properties. Moving on to Section \ref{sec4}, we present the fractional version of the CCE. Section \ref{sec5} is dedicated to discussing the empirical version. In Section \ref{sec7} a new distance measure between two copulas based on the Kullback-Leibler divergence and proposed a goodness of fit test procedure for copula based on it.  In Section \ref{sec6}, we conduct a Monte Carlo simulation study, to evaluate the $95$th percentile of the proposed test statistic. Moreover, real dataset is analyzed to illustrate the selection criteria of an appropriate copula based on the proposed distance measure. Finally, conclusion of the paper are discussed in Section \ref{cocl}. 
\section{Multivariate Cumulative Copula Entropy}\label{sec2}
In this section, we propose a $k$-dimensional CCE, which extend the bivariate CCE proposed by \cite{sunoj2023survival}. Let $C(\mathbf{u})$ be a $k$-dimensional copula, then $k$-dimensional CCE is defined as
\begin{equation*}
   \zeta\left(C\right)=-\int_{\mathbb{I}^k}C(\mathbf{u})\ln (C(\mathbf{u}))d\mathbf{u},
\end{equation*}
where $\mathbf{u}=(u_1,u_2,\dots,u_k)$. Since $f(x)=-x\ln(x)$ is non-negative and bounded by $e^{-1}$ on $\mathbb{I}$, it follows that $0\leq\zeta(C)\leq e^{-1}$. Now we consider some examples of the multivariate CCE of some well known multivariate copulas.

\begin{example}
    Consider the product copula $\Pi(\mathbf{u})=u_1u_2\dots u_k$, corresponds for the independence of random variables. Then, the $k$-dimensional CCE is given by
   $$ \zeta\left(\Pi\right)=\dfrac{k}{2^{k+1}},$$
   which is a decreasing function of $k \ (\geq 2)$. This implies that the uncertainty in a system of independent components decreases with increase in number of components.
\end{example}

\begin{example}
Consider the minimum copula $M({\bf{u}})=\min\{u_1,u_2,\dots, u_k\}$, then 
    \begin{align}\label{eq1}
       \zeta(M)=&-\int_{0}^{1}\int_{0}^{1} \cdots\int_{0}^{1} \min\{u_1,u_2,\dots, u_k\}\ln \left(\min\{u_1,u_2,\dots, u_k\}\right)du_1du_2\dots du_k
    \end{align}
    We can solve the above integral using the concept of order statistics. Let $U_1,U_2,\dots,U_k$ be $k$ random samples from uniform distribution over $\mathbb{I}$ and let $U_{(1)}=\min\{U_1,U_2,\dots,U_k\}$. The probability density function corresponds to $U_{(1)}$ is given by
   \begin{equation}\label{order}
       f_{U_{(1)}}(u)=\begin{cases} 
k(1-u)^{k-1} & \text{if } u\in\mathbb{I}, \\
0 & \text{otherwise.} 
\end{cases}
   \end{equation}
Now, the integral in Eq. (\ref{eq1}) can be viewed as $E(-U_{(1)}\ln \left(U_{(1)}\right),$ which is given by
\begin{align*}
   \zeta(M)=& E(-U_{(1)}\ln \left(U_{(1)}\right)\\
   =&- \int_0^1 u\ln (u)k(1-u)^{k-1}du\\
   =&-k\sum_{x=0}^{k-1}\binom{k-1}{x}(-1)^{x}\int_0^1 u^{x+1}\ln (u) du
   \\=&k\sum_{x=0}^{k-1}\binom{k-1}{x}\dfrac{(-1)^{x}}{(x+2)^2}.
\end{align*}
\end{example}
\begin{example}
    Consider the $k-$variate version of Cuadras-Aug{\'e} copula, proposed by \cite{cuadras2009constructing}, is given by \begin{equation}\label{cu}
C(\mathbf{u})=u_{(1)}\prod_{i=2}^ku_{(i)}^{\prod_{j=1}^{i-1}(1-\alpha_{ij})},
    \end{equation}
    where $u_{(1)}\leq u_{(2)}\cdots\leq u_{(k)}$ and $\alpha_{ij}\in\mathbb{I}.$ Let $\theta_1 = 1$, $\theta_i = \prod_{j=1}^{i-1}(1 - \alpha_{ij})$, and $p(i) = p(i-1) + \theta_i + 1$ with $p(1) = 2$, for $i = 2, 3, \dots, k$.
The CCE corresponds to Cuadras-Aug{\'e} copula is given by
    \begin{align*}
      \zeta(C)=&  -\int_{0}^{1}\int_{0}^{1} \cdots\int_{0}^{1} \prod_{i=1}^ku_{(i)}^{\theta_i} \ln\left(\prod_{i=1}^ku_{(i)}^{\theta_i}\right) \ du_1du_2\dots du_k\\
      =&-k!\int_{0}^{1}\int_{0}^{u_k} \int_{0}^{u_{k-1}} \cdots\int_{0}^{u_{2}} \prod_{i=1}^ku_{i}^{\theta_i} \ln\left(\prod_{i=1}^ku_{i}^{\theta_i}\right)du_1du_2\dots du_k\\
      =&k!\sum_{j=1}^k\theta_jI_j,
      \intertext{ where for every $j=1,2,3\dots,k$,}
      I_j=&-\int_{0}^{1}\int_{0}^{u_k} \int_{0}^{u_{k-1}} \cdots\int_{0}^{u_{2}}u_1u_2^{\theta_2}u_3^{\theta_3}\dots u_k^{\theta_k}\ln(u_j)du_1du_2\dots du_k\\
      =&\dfrac{1}{\prod_{i=1}^kp(i)}\left(\sum_{i=j}^k\dfrac{1}{p(j)}\right).
    \end{align*}
\end{example}
\par In literature, there exists several dependence measures for quantifying the dependence ability captured by the copula. One of the popular measure is Spearman's correlation. For bivariate case the Spearman's Rho for the copula $C$ is defined as
$$\rho_2(C)=12\int_{0}^{1}\int_{0}^{1}C(u_1,u_2)du_1du_2-3.$$
Due to the lack of symmetry, the concordance measures in the multivariate case, Spearman's Rho can be defined in two ways,
\begin{align}
   \rho_k^{-}(C)=&n(k)\left[2^k\int_{\mathbb{I}^k}C(\mathbf{u})d\mathbf{u}-1\right]\label{spm1}
   \intertext{and}
   \rho_k^{+}(C)=&n(k)\left[2^k\int_{\mathbb{I}^k}\pi(\mathbf{u})dC(\mathbf{u})-1\right],\nonumber
\end{align}
where $n(k)=\dfrac{k+1}{2^k-k-1}$ (for more details see \cite{schmid2010copula} and \cite{bedHo2016multivariate}). Using the multivariate version of Spearman's $\rho_k^{-}(C)$, we have the following theorem.
\begin{theorem}
    For every $k$-dimensional copula, $$\zeta(C)\leq-\mathcal{B}_k(C)\ln\left(\mathcal{B}_k(C)\right),$$
    where $\mathcal{B}_k(C)=2^{-k}\left[\dfrac{\rho_k^{-}(C)}{n(k)}+1\right]$ and $\rho_k^{-}(C)$ is the multivariate version of Spearman's correlation defined in Eq. (\ref{spm1}). 
\end{theorem}
\begin{proof}
  Using log-sum inequality, we have
  \begin{align*}
      \int_{\mathbb{I}^k}C(\mathbf{u})\ln \left(C(\mathbf{u})\right)d\mathbf{u}\geq&\left[\int_{\mathbb{I}^k}C(\mathbf{u})d\mathbf{u}\right]\left[\ln\left(\dfrac{\int_{\mathbb{I}^k}C(\mathbf{u})d\mathbf{u}}{\int_{\mathbb{I}^k}d\mathbf{u}}\right)\right]\\
      =&\mathcal{B}_k(C)\ln\left(\mathcal{B}_k(C)\right).
  \end{align*}
 The theorem follows by multiplying both sides by $-1$.
\end{proof}
\begin{definition}
    Let $C_1(\mathbf{u}), C_2(\mathbf{u}),\dots,C_m(\mathbf{u})$ be $m$ copulas of same dimension. Then, the weighted arithmetic mean of $m$ copulas is defined as $$C(\mathbf{u})=\displaystyle\sum_{i=1}^m\alpha_iC_i(\mathbf{u}),$$
where $\alpha_i\in \mathbb{I}, \ i=1,2,\dots,m$ with $\displaystyle\sum_{i=1}^m\alpha_i=1$.
\end{definition}
Note that the weighted arithmetic mean of $m$ copulas of same dimension is always a valid copula. 
\begin{theorem}
The  CCE of the weighted arithmetic mean of $m$ copulas never exceeds 
the weighted arithmetic mean of the CCEs of $m$ copulas of same dimension.
\end{theorem}
\begin{proof}
Let $C_1,C_2,\dots C_m$ be $m$ copulas and $C(\mathbf{u})=\sum_{i=1}^m\alpha_iC_i(\mathbf{u})$ be the arithmetic mean of $m$ copulas, where $\alpha_i\in \mathbb{I}, \ i=1,2,\dots,m$ and $\sum_{i=1}^m\alpha_i=1$. Since $f(x) = -x\ln(x)$ is concave on $\mathbb{I}$, it follows that for every $\alpha_i \in \mathbb{I}, \ i=1,2,\dots,m$, with $\sum_{i=1}^m\alpha_i = 1$, we have
   \begin{equation}\label{amc}
       f\left(\sum_{i=1}^m\alpha_ix_i\right)\leq\sum_{i=1}^m\alpha_if(x_i),
   \end{equation}
   for every $x_i\in\mathbb{I}$. Substituting $x_i=C_i(\mathbf{u})$ and integrating over $\mathbb{I}^k$, the result immediately follows.
\end{proof}
\begin{theorem}
    Let $\{C_n:n\in\mathbb{N}\}$ be a sequence of copulas of same dimension converges point-wise to $C$, then $\zeta(C_n)$ converges uniformly to $\zeta(C)$. 
\end{theorem}
\begin{proof}
  The sequence of copula $\{C_n:n\in\mathbb{N}\}$ converges point-wise to the copula $C$ implies that $C_n$ converges uniformly to $C$ (see Theorem 1.7.6 of \cite{durante2016principles}). It follows that for a given $\delta>0$, there exists $n_0\in\mathbb{N}$ such that
\begin{equation}\label{unifc}
    \left|C_n(\mathbf{u}) - C(\mathbf{u})\right| < \delta, \quad \text{for every $n \geq n_0$ and for every $\mathbf{u} \in \mathbb{I}^k$.} 
\end{equation}
Since $f(x)=-x\ln(x)$ is uniformly continuous on $\mathbb{I}$, it implies that for a given $\varepsilon>0$, there exists a $\delta>0$ such that \begin{equation}\label{unifc1}
    \left|f(x_1)-f(x_2)\right|<\varepsilon,
\end{equation}
whenever $|x_1-x_2|<\delta$. Substituting $x_1=C_n(\mathbf{u})$ and $x_2=C(\mathbf{u})$ in Eq.(\ref{unifc1}) and using Eq.(\ref{unifc}), we obtain 
$$\lim_{n\rightarrow\infty}-C_n(\mathbf{u})\ln (C_n(\mathbf{u}))=-C(\mathbf{u})\ln (C(\mathbf{u})).$$
Since $-C_n(\mathbf{u})\ln \left(C_n(\mathbf{u})\right)$ is bounded on $\mathbb{I}^k$, using bounded convergence theorem, we have
\begin{align*}
   \lim_{n\rightarrow\infty} \zeta(C_n)=&\lim_{n\rightarrow\infty}\int_{\mathbb{I}^k}-C_n(\mathbf{u})\ln \left(C_n(\mathbf{u})\right)d\mathbf{u}\\
   =&\int_{\mathbb{I}^k}\lim_{n\rightarrow\infty}-C_n(\mathbf{u})\ln \left(C_n(\mathbf{u})\right)d\mathbf{u}\\
   =&\zeta(C).
\end{align*}
\end{proof}
Now we will consider ordering property of the copula. 
\begin{definition} (Nelsen \cite{nelsen2007introduction})
    Let $C_1(\mathbf{u})$ and $C_2(\mathbf{u})$ be two $k$-dimensional copulas, then $C_1$ is said to be smaller than $C_2$ in concordance ordering, denoted by $C_1\prec C_2$, if $C_1(\mathbf{u})\leq C_2(\mathbf{u})$ for every $\mathbf{u}\in\mathbb{I}^k$.
\end{definition}
\begin{definition} 
    Let $C_1(\mathbf{u})$ and $C_2(\mathbf{u})$ be two $k$-dimensional copulas, then $C_1$ is said to be smaller than $C_2$ in entropy ordering, denoted by $C_1\prec_{E} C_2$, if $\zeta(C_1) \leq \zeta(C_2)$.
\end{definition}
\begin{remark}
    $C_1 \prec C_2 \nRightarrow C_1\prec_{E} C_2$, and conversely. 
\end{remark}
For counter-example, take $C_1(u_1,u_2)=\left(1+\left[\left(u^{-1}-1\right)^{2}+\left(v^{-1}-1\right)^{2}\right]^{0.5}\right)^{-1}$ and $C_2(u_1,u_2)=\min\{u_1,u_2\}$. It is a well-known result that $C_1\prec C_2$. But, $\zeta(C_1)=0.2790 $ and $\zeta(C_2)=0.2777$.
\begin{definition} (Shaked and Shanthikumar \cite{shaked2007stochastic})
    Let $X$ and $Y$ be two non-negative continuous random variables, each characterized by the CDF's $F$ and $G$, respectively. Let $f$ and $g$ denote the PDF corresponds to the CDF $F$ and $G$, respectively. Then, $X$ is said to be smaller in dispersive order, denoted by $F\leq_d G$, if and only if $f(F^{-1}(u))\geq g(G^{-1}(u))$, for all $u\in\mathbb{I}$. 
\end{definition}
\begin{definition}(Shaked and Shanthikumar \cite{shaked2007stochastic})
  Let $X$ and $Y$ be two non-negative continuous random variable with CDF's $F$ and $G$ and reversed hazard rates $r_X(t)$ and  $r_Y(t)$, respectively. Then $X$ is said to be smaller than $Y$ in reversed hazard rate ordering, denoted by $F\leq_{rh} G$, if and only if $r_X(t)\leq r_Y(t), \ \forall t>0$.
\end{definition}
It is to be noted that $r_X(t)\leq r_Y(t),\ \forall t>0$ if and only if $\dfrac{G(t)}{F(t)}$ is an increasing function of $t$. Bartoszewicz \cite{bartoszewicz1997dispersive} showed that if $F\leq_{rh} G$ and $X$ or $Y$ is increasing reversed failure rate (IRFR), then $G\leq_d F$. For a distribution $F$ is said to have IRFR if $\ln(F(x))$ is convex.
\par Every $k$-dimensional copula is nothing but a joint CDF of a multivariate random vector with marginals are distributed uniformly on $\mathbb{I}$. That is, $C(\mathbf{u})=P(U_1\leq u_1,U_2\leq u_2,\dots, U_k\leq u_k),$ where each $U_i$ follows uniform distribution over $\mathbb{I}$. Using the conditional distribution approach, we can represent every $k$-dimensional copula as the product of conditional distributions, which is discussed as follows.
\begin{align}
    C(\mathbf{u})=&P(U_1\leq u_1,U_2\leq u_2, U_3\leq u_3,\dots, U_k\leq u_k) \nonumber\\
    %=&P(U_1\leq u_1|U_2\leq u_2,U_3\leq u_3,\dots, U_k\leq u_k)P(U_2\leq u_2|U_3\leq u_3,\dots, U_k\leq u_k)\dots P(U_{k-1}\leq u_{k-1}|U_k\leq u_k)u_k\nonumber\\
    =&\prod_{i=1}^kC(u_i:u_k),\label{crep}
\end{align}
where $C(u_i:u_k)=P(U_i\leq u_i|U_{i+1}\leq u_{i+1},\dots U_k\leq u_k)=\dfrac{C(1,1,\dots,1,u_i,u_{i+1},\dots,u_k)}{C(1,1,\dots,1,u_{i+1},u_{i+2},\dots,u_k)}$ with $C(u_k:u_k)=u_k$.
The following theorem provides the condition for the entropy ordering of two copulas of same dimension.
\begin{theorem}
  Let $C_1$ and $C_2$ be two $k$-dimensional copulas. If $h(u_1,u_2,\dots,u_k)=\dfrac{C_1(u_1,u_2,\dots,u_k)}{C_2(u_1,u_2,\dots,u_k)}$ is an increasing function in each component $u_i$ ($i=1,2,\dots,k$) and other components are fixed, then the following statements are true.
    \begin{enumerate}[(a)]
        \item If $C_1(u_i:u_k)\leq_{d} C_2(u_i:u_k)$ for every $i=1,2,\dots,k$, then $C_1\prec_{E} C_2$.
        \item If $C_1(u_i:u_k)$ or $C_2(u_i:u_k)$  is IRFR for every $i=1,2,\dots,k$, then $C_1\prec_{E} C_2$.
    \end{enumerate}
\end{theorem}
\begin{proof}
   We will prove only the first part of the statement; the second part is the immediate consequence of the first part and is omitted. If $h(u_1,u_2,\dots,u_k)$ is an increasing function for each component  $u_i$, $i=1,2,\dots,k$, then it is easy to show that $C_1(u_i:u_k)\leq C_2(u_i:u_k)$, for every $\mathbf{u}\in\mathbb{I}^k$. Therefore,
   \begin{align*}
       \zeta(C_1)-\zeta(C_2)=&\int_{\mathbb{I}^k}\left(C_2(\mathbf{u})\ln\left(C_2(\mathbf{u})\right)-C_1(\mathbf{u})\ln\left(C_1(\mathbf{u})\right)\right)d\mathbf{u}\\
=&\sum_{i=1}^{k}\int_{\mathbb{I}^k}\left(\left[\prod_{j=1}^kC_2(u_j:u_k)\right]\ln\left(C_2(u_i:u_k)\right)-\left[\prod_{j=1}^kC_1(u_j:u_k)\right]\ln\left(C_1(u_i:u_k)\right)\right)d\mathbf{u}\\
&\hspace{12cm}\text{(using Eq. (\ref{crep}))}\\
\leq&\sum_{i=1}^{k}\int_{\mathbb{I}^k}\left[\prod_{\substack{j=1 \\ j \neq i}}^kC_2(u_j:u_k)\right]C_2(u_i:u_k)\ln\left(C_2(u_i:u_k)\right)d\mathbf{u}\\
&\hspace{2cm}-\sum_{i=1}^{k}\int_{\mathbb{I}^k}\left[\prod_{\substack{j=1 \\ j \neq i}}^kC_2(u_j:u_k)\right]C_1(u_i:u_k)\ln\left(C_1(u_i:u_k)\right)d\mathbf{u}.\\
=&\sum_{i=1}^{k}\int_{\mathbb{I}^k}\left[\prod_{\substack{j=1 \\ j \neq i}}^kC_2(u_j:u_k)\right]u_i\ln\left(u_i\right)\left(\dfrac{1}{f_2(C_2^{-1}(u_i:u_k))}-\dfrac{1}{f_1(C_1^{-1}(u_i:u_k))}\right)d\mathbf{u},
   \end{align*}
   where the last equation is obtained by substituting $t=C_j^{-1}(ui:u_k)$ for $j=1,2,$ and $f_j(\cdot)$ is the PDF corresponds to the CDF $C_j(u_i:u_k)$, and $C_j^{-1}(u_i:u_k)$ is the inverse of $C_j(u_i:u_k)$, $j\in\{1,2\}$. The result follows by using the definition of dispersive ordering.
\end{proof}

 \section{Cumulative Copula Information Generating Function}\label{sec3}
In this section, we introduce a generating function for CCE and study its important properties. Let $C(\mathbf{u})$ be a $k$- dimensional copula, then we define the cumulative copula information generating function (CCIGF) as follows.
\begin{definition}
    Let $C(\mathbf{u})$ be a $k$-dimensional copula, then CCIGF, defined as \begin{equation}\label{cg}
\mathcal{G}_C(s)=\int_{\mathbb{I}^k}\left[C(\mathbf{u})\right]^sd\mathbf{u}, \ s>0.
    \end{equation}
\end{definition}
It is easy to show that the first derivative of $\mathcal{G}_C(s)$ at $s=1$ reduces to $-\zeta(C)$. So, we call $\mathcal{G}_C(s)$ as information generating function. Moreover, $\mathcal{G}_C(1)=\mathcal{B}_k(C)=2^{-k}\left[\dfrac{\rho_k^{-}(C)}{n(k)}+1\right]$ and $\mathcal{G}_C(2)=4J(C)$,
where $J(C)=\dfrac{1}{4}\int_{\mathbb{I}^k}C(\mathbf{u})d \mathbf{u}$ is cumulative copula extropy proposed by \cite{saha2023copula}. The CCIGFs of some well known copulas are provided in Table \ref{cceg}.
\begin{table}[H]
\centering
\caption{CCIGF of some well known copulas}\label{cceg}
\scalebox{0.6}{
\begin{tabular}{ll}
\hline
Copula                                                                                                                                                                                         &  $\mathcal{G}_C(s)$                                                                                                                            \\ \hline 
FGM Copula: $C(u_1,u_2)=u_1u_2\left(1+\theta(1-u_1)(1-u_2)\right),\theta\in\mathbb{I}.$                                                                                                        & $\displaystyle\sum_{x=0}^{\infty}\binom{s+x-1}{x}\theta^x\left[\beta(s+1,x+1)\right]^2$, where $\beta(p,q)$ is the usual beta function,                                                                                              \\
Marshall Olkin Copula: $C(u_1,u_2)=u_1^{\alpha_1}u_2^{\alpha_2}\min\{u_1^{\alpha_1},u_2^{\alpha_2}\},\alpha_1,\alpha_2\in\mathbb{I}$                                                           & $\dfrac{1}{(s+1)}\left[\dfrac{\alpha_1}{\alpha_2(s+1)+\alpha_1s(1-\alpha_2)}+\dfrac{\alpha_2}{\alpha_1(s+1)+\alpha_2s(1-\alpha_1)}\right]$                     \vspace{.25cm}            \\
  
Product Copula: $\Pi(\mathbf\mathbf{u})=u_1u_2\dots u_k$                                                                                                                                 & $\displaystyle(s+1)^{-k}$                                                                                                                                                            \\
Cuadras-Aug{\'e} Copula : $C(\mathbf{u})=\displaystyle u_{(1)}\prod_{i=2}^ku_{(i)}^{\prod_{j=1}^{i-1}(1-\alpha_{ij})}, \alpha_{ij}\in\mathbb{I}$ & $k!\displaystyle \prod_{i=1}^k[q(i)]^{-1}, q(i)=\theta_is+1+q(i-1)$, $q(1)=s+1,\theta_1 = 1,\theta_i = \prod_{j=1}^{i-1}(1 - \alpha_{ij}),i\geq2$ \\ \hline
\end{tabular}}
\end{table}
  Using the definition of multivariate Spearman's Rho defined in Eq. (\ref{spm1}), we have the following theorem. 
\begin{theorem}\label{prop4.1}
    For any $k$-dimensional copula $C$, the following inequality holds.
    $$\mathcal{G}_C(s)  \begin{cases} 
\geq \left[\mathcal{B}_k(C)\right]^s, & \text{if } \  s > 1 \\
\leq \left[\mathcal{B}_k(C)\right]^s, & \text{if} \hspace{0.2cm} 0\leq s\leq 1.
\end{cases}$$
\end{theorem}
\begin{proof}
    For $s\in\mathbb{I}$, $f(x)=x^s$ is concave on $x\in\mathbb{I}$ and convex on $x\in\mathbb{I}$ for $s>1$. Using Jensens's inequality and using definition of multivariate Spearman's Rho defined in Eq. (\ref{spm1}), theorem immediately follows. 
\end{proof}
The following theorem discusses about the ordering property of the CCIGF. The proof is straightforward, so omitted here. 
\begin{theorem}\label{prop4.2}
    Let $C_1$ and $C_2$ be two copula of same dimension, then if $C_1\prec C_2$, then $\mathcal{G}_{C_1}(s)\leq\mathcal{G}_{C_2}(s)$, for every $s>0$.
\end{theorem}
The following theorem, is due to Theorem \ref{prop4.2}, which provides a tight bounds for every .
\begin{theorem}
   For any $k$-dimensional copula $C$, $$\left(\displaystyle\prod_{i=1}^{k}(s+i)\right)^{-1}\leq \mathcal{G}_C(s) \leq k\beta(s+1,k),$$ 
   where $\beta(p,q)$ is the usual Beta function and $s>0$.
\end{theorem}
\begin{proof}
    From Eq. (\ref{cg}), we have
    \begin{align*}
      \mathcal{G}_C(s)=& \int_{\mathbb{I}^k}\left[C(\mathbf{u})\right]^sd\mathbf{u} \\
      \geq& \int_{\mathbb{I}^k}\left[W(\mathbf{u})\right]^sd\mathbf{u}\\
      =& \underbrace{\int_{0}^{1}\int_{0}^{1} \cdots\int_{0}^{1}}_{k \text{ times}}  \left(\max\left\{\displaystyle \sum_{i=1}^{k}u_i-k+1,0\right\}\right)^s du_1du_2\dots du_k\\
%=& \int_{0}^{1}\int_{1-u_1}^{1} \int_{2-u_1-u_2}^{1}\cdots\int_{k-1-\sum_{i=1}^{k-1}u_i}^{1}  \left(\displaystyle \sum_{i=1}^{k}u_i-k+1\right)^sdu_1du_2\dots du_k\\
%=&\int_{0}^{1}\int_{0}^{t_1} \int_{0}^{t_2}\cdots\int_{0}^{t_{k-1}}  t_k^s \ dt_kdt_{k-1}\dots dt_1\\
      =&\left(\displaystyle\prod_{i=1}^{k}(s+i)\right)^{-1}.
    \end{align*}
Similarly, using the upper bound of every copula, we have   
\begin{align*}
    \mathcal{G}_C(s)\leq&\int_{\mathbb{I}^k}\left[M(\mathbf{u})\right]^sd\mathbf{u}\\
      =& \underbrace{\int_{0}^{1}\int_{0}^{1} \cdots\int_{0}^{1}}_{k \text{ times}}  \left(\min\left\{u_1,u_2,\dots,u_k\right\}\right)^s du_1du_2\dots du_k\\
      =& k\int_{0}^{1}u^s(1-u)^{k-1}du\\
      =&k\beta(s+1,k).
\end{align*}
\end{proof}
\begin{definition} 
   Let $C_1, C_2, \dots, C_m$ be $m$ copulas of the same dimension. The weighted geometric mean of $m$ copulas is defined as 
\[ C(\mathbf{u}) = C_1(\mathbf{u})^{\alpha_1}C_2(\mathbf{u})^{\alpha_2} \dots C_m(\mathbf{u})^{\alpha_m}, \]
where $\alpha_i \in \mathbb{I}$ for $i = 1, 2, \dots, m$ with $\displaystyle \sum_{i=1}^{m}\alpha_i = 1$.
\end{definition}
\begin{remark}

The weighted geometric mean of copulas may not always be a valid copula. However, it can be a valid copula under certain conditions. For further details, one may refer to \cite{cuadras2009constructing} and \cite{zhang2013some}.
\end{remark}
\begin{theorem}
    The CCIGF of the weighted geometric mean of $m$ copulas never exceeds the weighted geometric mean of s of $m$ copulas. 
\end{theorem}
\begin{proof}
   Let $C_1, C_2, \dots, C_m$ be $m$ copulas, and $C(\mathbf{u}) = C_1(\mathbf{u})^{\alpha_1}C_2(\mathbf{u})^{\alpha_2} \dots C_m(\mathbf{u})^{\alpha_m}$ be the weighted geometric mean (WGM) of $m$ copulas, where $\alpha_i \in \mathbb{I}$ for $i=1,2,\dots,m$ and $\displaystyle\sum_{i=1}^m\alpha_i=1$. Let $\mathcal{G}_C(s)$ and $\mathcal{G}_{C_i}(s);i=1,2,\dots,m$ be the s of $C(\mathbf{u})$ and $C_i(\mathbf{u});i=1,2,\dots,m$. The information generating function of the WGM of $m$ copulas is given by
\begin{align*}
    \mathcal{G}_C(s)=&\int_{\mathbb{I}^k}\left[C(\mathbf{u})\right]^sd\mathbf{u}\\
=&\int_{\mathbb{I}^k}\left[C_1(\mathbf{u})^{\alpha_1}C_2(\mathbf{u})^{\alpha_2} \dots. C_m(\mathbf{u})^{\alpha_m}\right]^sd\mathbf{u} 
  \end{align*}
  Using H\"older's inequality on $m$ integrals (see \cite{kufner1977function}, \cite{finner1992generalization}), we have
\begin{align*}
\mathcal{G}_C(s)\leq \left(\int_{\mathbb{I}^k}\left[C_1(\mathbf{u})\right]^sd\mathbf{u}\right)^{\alpha_1}\left(\int_{\mathbb{I}^k}\left[C_2(\mathbf{u})\right]^sd\mathbf{u}\right)^{\alpha_2}\dots\left(\int_{\mathbb{I}^k}\left[C_m(\mathbf{u})\right]^sd\mathbf{u}\right)^{\alpha_m}
=\prod_{i=1}^{m}\left(\mathcal{G}_{C_1}(s)\right)^{\alpha_i}.
\end{align*}
\end{proof}
The proofs of the following theorems are similar to the proofs given in Section \ref{sec2}, so we left out here.
\begin{theorem}
Let $C_1,C_2,\dots C_m$ be $m$ copulas and $C(\mathbf{u})=\displaystyle\sum_{i=1}^m\alpha_iC_i(\mathbf{u})$ be the arithmetic mean of $m$ copulas, where $\alpha_i\in \mathbb{I}, \ i=1,2,\dots,m$ and $\displaystyle\sum_{i=1}^m\alpha_i=1$. Let $\mathcal{G}_{C}(s)$ and $\mathcal{G}_{C_i}(s);i=1,2,\dots,m$ be the  of $C$ and $C_i;i=1,2,\dots,m$, then
$$\mathcal{G}_C(s)  \begin{cases} 
\geq \displaystyle\sum_{i=1}^m\alpha_i\mathcal{G}_{C_i}(s), & \text{if } \  s > 1 \\
\leq \displaystyle\sum_{i=1}^m\alpha_i\mathcal{G}_{C_i}(s), & \text{if} \hspace{0.2cm} 0\leq s\leq 1.
\end{cases}$$
\end{theorem}
\begin{theorem}
    Let $\{C_n:n\in\mathbb{N}\}$ be a sequence of copulas of same dimension converges point-wise to $C$, then $\mathcal{G}_{C_n}(s)$ converges uniformly to $\mathcal{G}_C(s)$, for every $s>0$. 
\end{theorem}
 \section{Fractional Multivariate Cumulative Copula Entropy}\label{sec4}
 In this section, we generalizes the concept of multivariate CCE using fractional calculus. Using Riemann-Liouville fractional derivative, Ubraico \cite{ubriaco2009entropies} proposed the fractional Shannon entropy which is given in Eq. (\ref{fse}). Xiong et al. \cite{xiong2019fractional} and Kayid and Shrahili \cite{kayid2022some} further extended the concept to propose the fractional version of cumulative residual entropy and cumulative entropy, respectively. To the best of our knowledge, no prior work has addressed the fractional version of copula entropy, even in the context of bivariate cases.
 \begin{definition}
     Let $C(\mathbf{u})$ be a $k$-dimensional copula, then fractional cumulative copula entropy (FCCE) can be defined as
     \begin{equation}\label{fcce}
\zeta_{r}(C)=\int_{\mathbb{I}^k}C(\mathbf{u})\left(-\ln (C(\mathbf{u})\right)^r d\mathbf{u},\ 0\leq r\leq1.
     \end{equation}
 \end{definition}
 For $r=1$, FCCE reduces to CCE. Following are examples of FCCE of some well-known bivariate and multivariate copulas.
 \begin{example}
The fractional CCE of Fr{\'e}chet-Hoeffding lower bound copula is given by
\begin{align*}
    \zeta_r(W)=&\int_0^1\int_0^1 \max\{u_1+u_2-1,0\}\left[-\ln\left(\max\{u_1+u_2-1,0\}\right)\right]^rdu_1du_2\\
    =&\int_0^1\int_{1-u_2}^1\left(u_1+u_2-1\right)\left[-\ln\left(u_1+u_2-1\right)\right]^rdu_1du_2\\
    =&\int_0^1\int_{0}^{u_2} u_1[-\ln(u_1)]^rdu_1du_2
    \end{align*}
    Using the transformation $t=-\ln(u_1)$, we get $\zeta_r(W)=\Gamma(r+1)\left(2^{-r-1}-3^{-r-1}\right).$
  %  =&\int_0^1\int_{0}^{u_2} u_1[-\ln(u_1)]^rdu_1du_2\\
  %  =&\int_0^1\int_{-\ln u_2}^\infty e^{-2t}t^rdtdu_2 \hspace{6.4cm}(\text{Using the transformation $t=-\ln(u_1)$.})\\
  %  =&\int_0^\infty\int_{e^{-t}}^1 e^{-2t}t^rdu_2dt\\
  %  =&\Gamma(r+1)\left(2^{-r-1}-3^{-r-1}\right).
%\end{align*}
 \end{example}
 \begin{example}
     Consider the bivariate Cuadras-Aug{\'e} copula given by
     $$C(u_1,u_2)=\left(\min\{u_1,u_2\}\right)^{1-\alpha}\left(u_1u_2\right)^{\alpha}, \ \alpha\in\mathbb{I}^2.$$
     Then the FCCE corresponds to Cuadras-Aug{\'e} copula is 
     \begin{align*}
         \zeta_r(C)=&\int_0^1\int_0^1 \left(\min\{u_1,u_2\}\right)^{1-\alpha}\left(u_1u_2\right)^{\alpha}\left[-\ln\left(\left(\min\{u_1,u_2\}\right)^{1-\alpha}\left(u_1u_2\right)^{\alpha}\right)\right]^rdu_1du_2\\
         =&\dfrac{2!\Gamma(r+1)}{1-\alpha}\left[\dfrac{1}{2^{r+1}}-\left(\dfrac{\alpha+1}{\alpha+3}\right)^{r+1}\right].
        % =&2!\int_0^1\int_0^{u_2} u_1u_2^{\alpha}\left[-\ln\left(u_1u_2^{\alpha}\right)\right]^rdu_1du_2\\
         %=&2!\int_O^{\infty}\int_{e^{-\frac{t}{\alpha+1}}}^1e^{-2t}t^ru_2^{-\alpha}du_2dt\hspace{6cm}(\text{Using the transformation $t=-\ln(u_1)$.})\\
        % =&\dfrac{2!\Gamma(r+1)}{1-\alpha}\left[\dfrac{1}{2^{r+1}}-\left(\dfrac{\alpha+1}{\alpha+3}\right)^{r+1}\right].
     \end{align*}
 \end{example}
\begin{example}
    The FCCE corresponds to the $k-$dimensional product copula is 
     \begin{align*}
         \zeta_r(\Pi)=\int_0^1\int_0^1\cdots\int_0^1u_1u_2\dots u_k\left[-\ln\left(u_1u_2\dots u_k\right)\right]^rdu_1du_2\dots du_k=\dfrac{\Gamma(r+k)}{2^{r+k}}.
        % =&\int_0^1\int_0^1\cdots\int_0^1\int_{-\ln\left(u_2u_3\dots u_k\right)}^{\infty}\dfrac{e^{-2t}t^r}{u_2u_3\dots u_k}du_1du_2\dots du_k\hspace{2.7cm}(\text{Using the transformation $t=-\ln(u_1)$.})\\
       %  =&\int_0^{\infty}\int_{e^{-t}}^{1}\int_{e^{-t}u_2^{-1}}^{u_2^{-1}}\cdots\int_{e^{-t}u_2^{-1}\dots u_k^{-1}}^{u_2^{-1}\dots u_k^{-1}}\dfrac{e^{-2t}t^r}{u_2u_3\dots u_k}du_kdu_{k-1}\dots dt\\
        % =&\dfrac{\Gamma(r+k)}{2^{r+k}}.
     \end{align*}
\end{example}
\begin{example}
    The FCCE corresponds to the minimum copula is 
     \begin{align*}
         \zeta_r(M)=&\int_0^1\int_0^1\cdots\int_0^1\min\left\{u_1,u_2,\dots, u_k\right\}\left[-\ln\left(\min\left\{u_1,u_2,\dots, u_k\right\}\right)\right]^rdu_1du_2\dots du_k\\
         =&k\sum_{x=0}^{k-1}\binom{k-1}{x}(-1)^x\dfrac{\Gamma(r+1)}{(x+2)^{r+2}}.
         %=&k\int_0^1u\left[-\ln(u)\right]^r(1-u)^{k-1} \quad (\text{Using Eq. (\ref{order}) })\\
        % =&k\sum_{x=0}^{k-1}\binom{k-1}{x}(-1)^x\int_0^{1} u^{x+1}\left[-\ln(u)\right]^rdu\\
        % =&k\sum_{x=0}^{k-1}\binom{k-1}{x}(-1)^x\int_0^{\infty} t^{r}e^{-(x+2)t}dt\hspace{5cm}(\text{Using the transformatiom $t=-\ln(u)$}.)\\
         %=&k\sum_{x=0}^{k-1}\binom{k-1}{x}(-1)^x\dfrac{\Gamma(r+1)}{(x+2)^{r+2}}.
     \end{align*}
\end{example}
In the following theorem, we obtain an upper bound for FCCE in terms of CCE.
 \begin{theorem}
     Let $C(\mathbf{u})$ be a $k$-dimensional copula, then $\left(\zeta(C)\right)^r\geq \zeta_{r}(C)$, for every $0\leq r\leq 1$. 
 \end{theorem}
 \begin{proof}
     For fixed $r\in\mathbb{I}$, the function $f(x)=x^r$ is concave on $x\in\mathbb{I}$. Using Jensen's inequality on concave function, we have
  \begin{align*}
  \left(\zeta(C)\right)^r=&\left(-\int_{\mathbb{I}^k} C(\mathbf{u}) \ln (C(\mathbf{u})) d\mathbf{u}\right)^r\\
  \geq &\int_{\mathbb{I}^k}\left(- C(\mathbf{u}) \ln (C(\mathbf{u}))\right)^r d\mathbf{u}\\
  \geq&\int_{\mathbb{I}^k} C(\mathbf{u}) \left(-\ln (C(\mathbf{u}))\right)^r d\mathbf{u} \\
  =&\zeta_{r}(C).
\end{align*}
 \end{proof}
The proofs of the following theorems are similar to the proofs given in the Section \ref{sec2}, so we omitted.
\begin{theorem}
The weighted arithmetic mean of the FCCEs of $m$ copulas of same dimension is always greater than the FCCE of the weighted arithmetic mean of $m$ copulas.
\end{theorem}
\begin{theorem}
    Let $\{C_n;n\in\mathbb{N}\}$ be a sequence of copulas of same dimension converges point-wise to $C$. Then, $\displaystyle\lim_{n\rightarrow\infty}\zeta_r(C_n)=\zeta_r(C)$, for every $r\in\mathbb{I}$. 
\end{theorem}
\begin{theorem}
  Let $C_1$ and $C_2$ be two $k$-dimensional copulas. If $h(u_1,u_2,\dots,u_k) = \frac{C_1(u_1, u_2, \dots, u_k)}{C_2(u_1, u_2, \dots, u_k)}$ is an increasing function for each component $u_i$, $i = 1, 2, \dots, k$ and others are fixed, then the following statements hold.
    \begin{enumerate}[(a)]
        \item If $C_1(u_i:u_k) \leq_{d} C_2(u_i:u_k)$ for every $i = 1, 2, \dots, k$, then $\zeta_r(C_1) \leq \zeta_r(C_2)$, for every $r \in \mathbb{I}$.
        
        \item If $C_1(u_i:u_k)$ or $C_2(u_i:u_k)$  is IRFR for every $i=1,2,\dots,k$, then $\zeta_r(C_1) \leq \zeta_r(C_2)$, for every $r \in \mathbb{I}$.
    \end{enumerate}
\end{theorem}

\section{Empirical Beta Cumulative Copula Entropy}\label{sec5}
In this section, we will develop a non-parametric estimator of CCE and its generating function using empirical beta copula. First, we will discuss the empirical copula which is defined as follows.
\begin{definition}(Nelsen \cite{nelsen2007introduction})
Let $\mathbf{X}_i,i=1,2,\dots,N$, be $N$ random samples from a $k$-variate distribution, where each $\mathbf{X}_i$ consists of $k$ components denoted by $\mathbf{X}_i=\left(X_{i,1}, X_{i,2},\dots,X_{i,k}\right)'$. Let $R_{i,j}$ be the rank of the observation $X_{i,j}$, for every $i=1,2,\dots,N$ and $j=1,2,\dots,k$. Then the empirical copula can be defined as
\[ C_N\left(\mathbf{u}\right) =\dfrac{1}{N}\sum_{i=1}^N\prod_{j=1}^{k}\mathbf{I}\left(\dfrac{R_{i,j}}{N+1}\leq u_{j}\right), \]
where $\mathbf{I}(\cdot)$ is the usual indicator function.

\end{definition}
Sunoj and Nair \cite{sunoj2023survival} proposed an empirical CCE for the bivariate case. But computationally evaluating the empirical CCE is a time-consuming task for higher dimensional case. Moreover, the empirical copula is not even a valid copula. \cite{segers2017empirical} proposed empirical beta copula given by 
\begin{equation*}
  \hat{C}_N(\mathbf{u})=\dfrac{1}{N}\sum_{i=1}^N\prod_{j=1}^{k}\sum_{y=R_{i,j}}^N\binom{N}{y}u_j^y(1-u_j)^{N-y},
\end{equation*}
where $R_{i,j}$ is the rank of the $i^{th}$ observation of the $j^{th}$ component $X_{i,j}$. It is to be noted that empirical beta copula is a valid copula when there is no ties in the data. Moreover, the empirical beta copula is a particular case of empirical Bernstein copula, introduced by \cite{sancetta2004bernstein}, when all Bernstein polynomials have degrees equal to the sample size $N$. In case of ties, we need to break the ties at random, then the empirical beta copula will become a valid copula. Furthermore, empirical beta copula provides better estimate compared to empirical copula in terms of bias and variance (see \cite{segers2017empirical}).   Using the definition of empirical beta copula, we can define the empirical beta CCE which is given below.
\begin{definition}
     Let $\mathbf{X}_j,j=1,2,\dots,N$ be $N$ random samples from a continuous $k$-variate distribution, then the empirical beta CCE can be defined as
     \begin{equation}\label{ecce}
         \zeta(\hat{C}_N)=-\int_{\mathbb{I}^k}\hat{C}_N(\mathbf{u})\ln\left(\hat{C}_N(\mathbf{u})\right)d\mathbf{u}.
     \end{equation}
\end{definition}
Analogous to empirical CCE defined in Eq. (\ref{ecce}), we can define the fractional version of empirical beta CCE and empirical beta .
\begin{definition}
    The fractional empirical beta CCE corresponds to the $N$ random samples is given by 
     \begin{equation*}
         \zeta_r(\hat{C}_N)=\int_{\mathbb{I}^k}\hat{C}_N(\mathbf{u})\left[-\ln\left(\hat{C}_N(\mathbf{u})\right)\right]^rd\mathbf{u}, r\in\mathbb{I}.
     \end{equation*}
\end{definition}
\begin{definition}
    For every $N$ random samples, the empirical beta  can be defined as
    \begin{equation*}
         \mathcal{G}_{\hat{C}_N}(s)=\int_{\mathbb{I}^k} \left[\hat{C}_N(\mathbf{u})\right]^s, s>0.
     \end{equation*}
\end{definition}
The following theorem asserts that the  fractional empirical beta CCE and the empirical beta  are always consistent estimators for the FCCE and its generating function.
\begin{theorem}
    The  fractional empirical  beta CCE and empirical beta  converges to the CCE and  almost surely. That is 
    \begin{enumerate}[(a)]\label{pr1}
        \item $\zeta_r(\hat{C}_N) \stackrel{a.s.}{\longrightarrow} \zeta_r(C)$, as $N\rightarrow\infty$ and for every $r\in\mathbb{I}$,
        \item $\mathcal{G}_{\hat{C}_N}(s) \stackrel{a.s.}{\longrightarrow} \mathcal{G}_C(s)$, as $N\rightarrow\infty$ and for every $s>0$.
    \end{enumerate}
\end{theorem}
\begin{proof}
We prove only the first part of the theorem, second  part is similar and are therefore omitted.
Let $\mathbf{X}_j$, for $j=1,2,\dots,N$, represent $N$ random samples from a continuous $k$-variate distribution with underlying copula $C$. Let $C_N$ be an empirical copula obtained from the sample. Then, Glivenco-Cantelli theorem on empirical copula states that
\begin{equation}\label{s1}
\sup_{\mathbf{u}\in\mathbb{I}^k}\left|C_N(\mathbf{u})-C(\mathbf{u})\right|\stackrel{a.s.}{\longrightarrow}0,
\end{equation}
    as $N\rightarrow\infty.$ For more details, one could refer \cite{ec1}, \cite{ec2}, \cite{ec3} and \cite{ec4}. 
\cite{segers2017empirical} showed that for any $k-$dimensional copula $C$,
\begin{equation}\label{s2}
  \sup_{\mathbf{u}\in\mathbb{I}^k}\left|C_N(\mathbf{u})-\hat{C}_N(\mathbf{u})\right|\leq k\left[\left(\dfrac{\ln N}{N}\right)^{1/2}+N^{-1/2}+N^{-1}\right]\stackrel{a.s.}{\longrightarrow}0,
\end{equation}
as $N\rightarrow\infty.$ Using Eq.(\ref{s1}) and Eq.(\ref{s2}), we have 
\begin{equation*}
  \sup_{\mathbf{u}\in\mathbb{I}^k}\left|\hat{C}_N(\mathbf{u})-C(\mathbf{u})\right|\stackrel{a.s.}{\longrightarrow}0,  
\end{equation*}
    as $N\rightarrow\infty.$ Since we $f(x)=-x\ln(x)$ is continuous on $\mathbb{I}$, it follows that 
$$\displaystyle\lim_{N\rightarrow\infty}\sup_{\mathbf{u}\in\mathbb{I}^k}\left|-\hat{C}_N(\mathbf{u})\ln(C_N(\mathbf{u}))+C(\mathbf{u})\ln(C(\mathbf{u}))\right|=0,$$
almost surely as $N\rightarrow\infty$.
Since the CCE is always bounded, using dominated convergence theorem the result immediately follows.
\end{proof}
Now, we will illustrate the consistency property of the fractional empirical beta CCE and empirical beta  by considering various bivariate and trivariate copulas available in the literature. We consider the following multivariate copulas for the illustration purpose:
\begin{enumerate}
\item Product Copula: $\Pi(\mathbf{u})=\Pi_{i=1}^ku_i$.
    \item Clayton Copula: $C(\mathbf{u})=\max\left\{\sum_{i=1}^ku_i^{\alpha}-k+1,0\right\}^{-1/\alpha}, \alpha\in[-1,\infty)\setminus\{0\}$.
    \item Gumbel-Hougaard Copula: $C(\mathbf{u})=\exp\left\{-\left(\sum_{i=1}^k(-\ln(u_i))^{\phi}\right)^{1/\phi}\right\}, \phi\geq 1$.
    \item Frank Copula: $C(\mathbf{u})=-\theta^{-1}\ln\left(1+\dfrac{\prod_{i=1}^ke^{-\theta u_i-1}}{e^{-\theta}-1}\right),\theta\in\mathbb{R}\setminus\{0\}.$
    \item Joe Copula: $C(\mathbf{u}) =  1 - \left(1 - \left[1 - (1 - u_1)^\theta\right]\cdots \left[1 - (1 - u_K)^\theta\right]\right)^{1/\theta}
$, where $\theta\geq1$.
    \item Normal copula: $C(\mathbf{u})=\mathbf{\Phi}_{\boldsymbol{\rho}}\left(\phi(u_1),\phi(u_2),\dots,\phi(u_k)\right),$
    where $\mathbf{\Phi}_{\boldsymbol{\rho}}(\cdot)$ is the CDF of multivariate normal distribution with zero mean and correlation matrix $\boldsymbol{\rho}=[\rho_{ij}]$ with each \(|\rho_{ij}| < 1\) and $\rho_{ij}=\rho_{ji}$ for \(i \neq j = 1, 2, 3, \ldots, k\).

\end{enumerate}
\par
\begin{figure}[H]
    \centering
    \subfloat[Clayton copula with parameter $\alpha=0.6$]{\includegraphics[height=5cm, width=5cm]{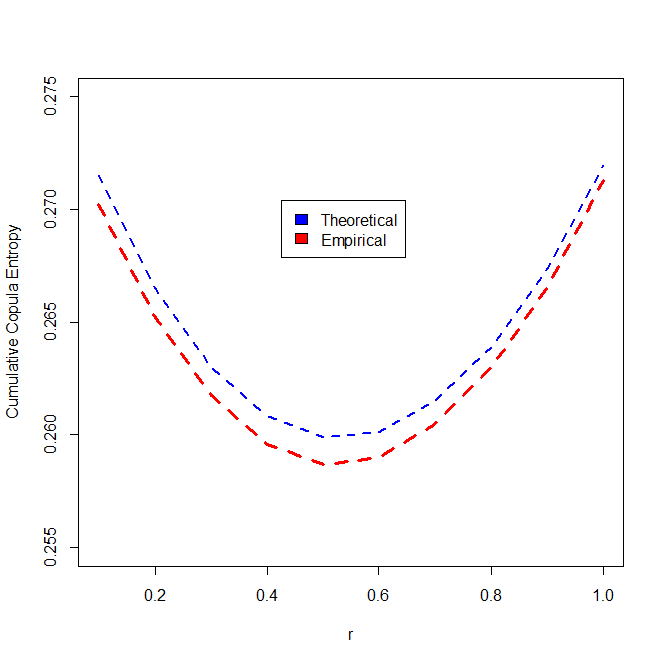}} \hfill
    \subfloat[Gumbel-Hougaard copula with parameter $\phi=2$]{\includegraphics[height=5cm, width=5cm]{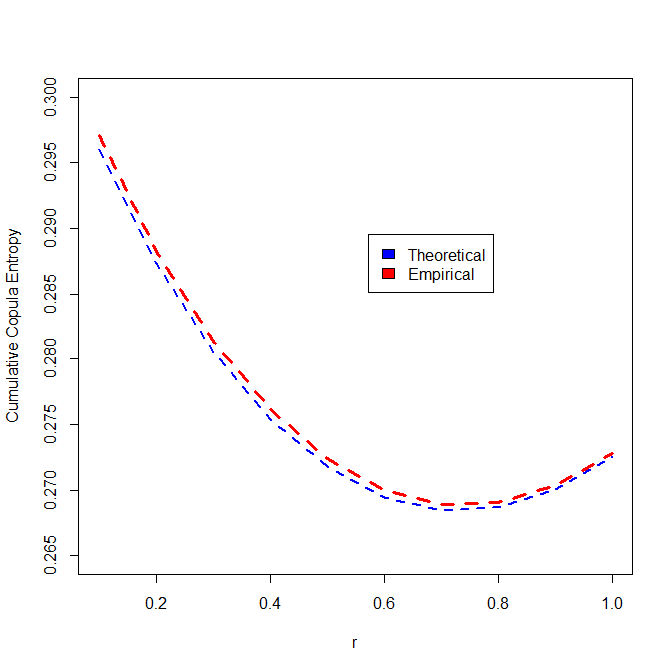}} \hfill
    \subfloat[Gaussian copula with parameters $\rho_1=0.2, \rho_2=0.6$]{\includegraphics[height=5cm, width=5cm]{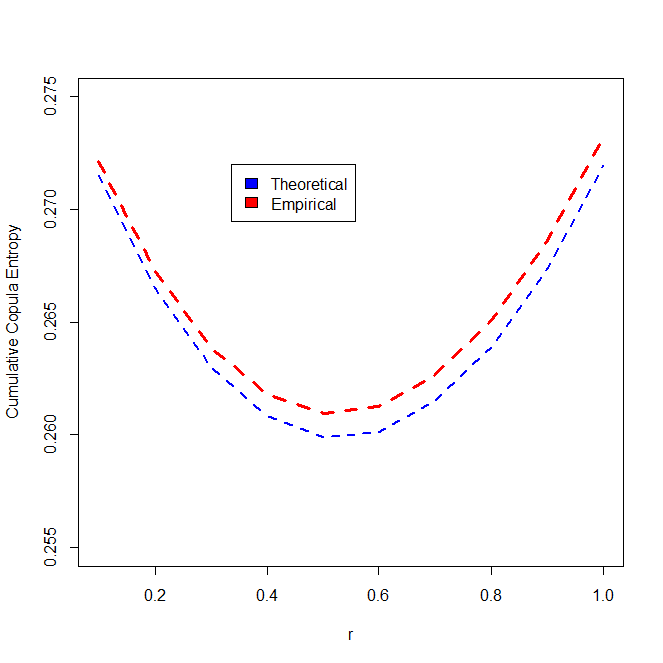}}
    \caption{The fractional empirical beta CCE and theoretical fractional CCE of various bivariate copulas.}\label{fig1}
\end{figure}

\begin{figure}[H]
    \centering
    \subfloat[Clayton copula with parameter $\alpha=0.6$]{\includegraphics[height=5cm, width=5cm]{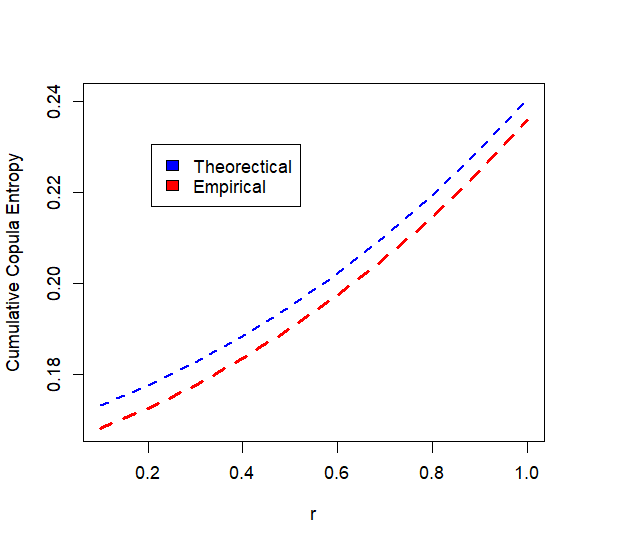}} \hfill
    \subfloat[Gumbel-Hougaard copula with parameter $\phi=2$]{\includegraphics[height=5cm, width=5cm]{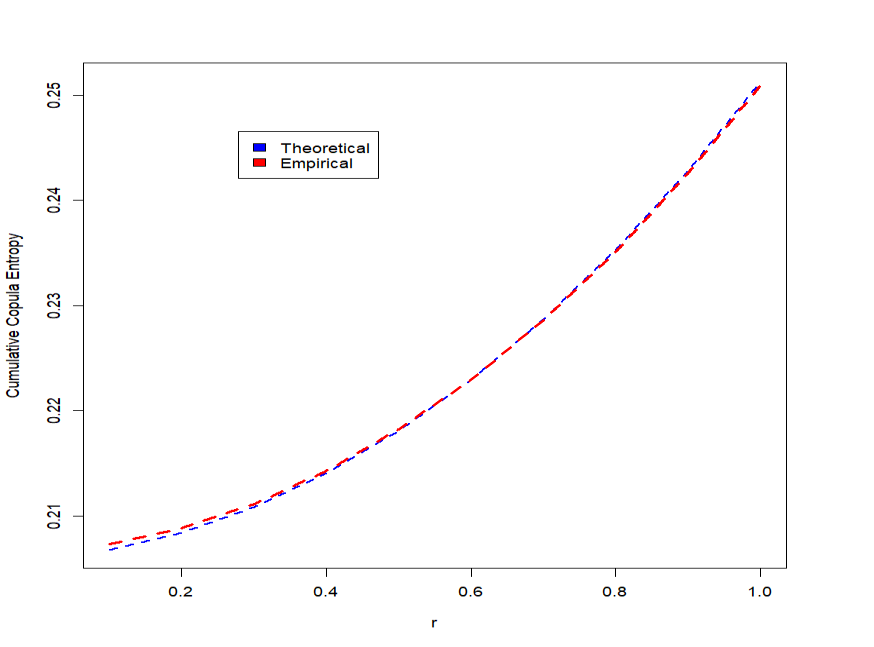}} \hfill
    \subfloat[Normal copula with parameters $\rho_1=0.2, \rho_2=0.6, \rho_3=0.9$]{\includegraphics[height=5cm, width=5cm]{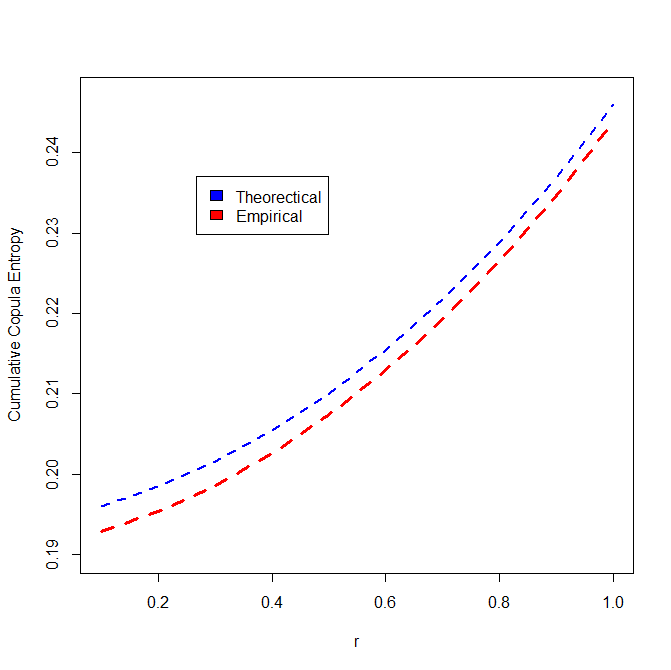}}
    \caption{The fractional empirical beta CCE and theoretical fractional CCE of various trivariate copulas.}\label{fig2}
\end{figure}

\begin{figure}[H]
    \centering
      \subfloat[Product Copula]{\includegraphics[height=5cm, width=5cm]{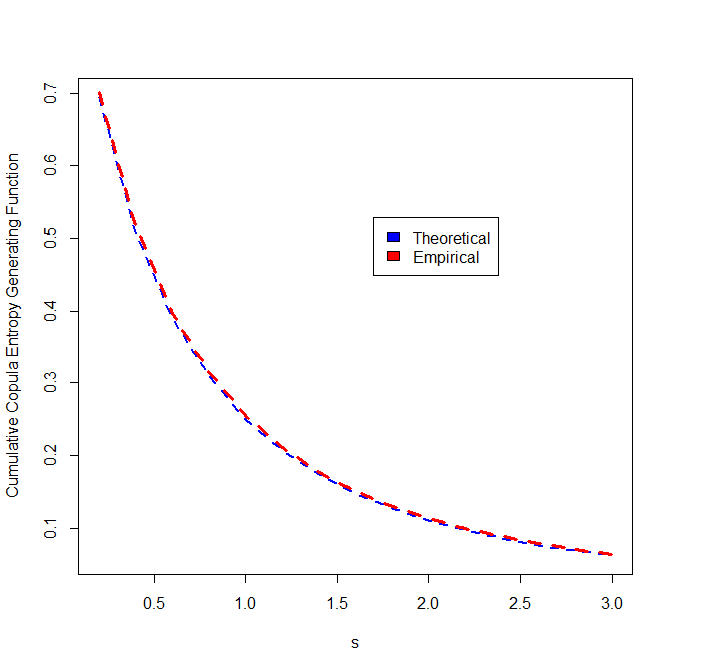}}\hfill
      \subfloat[Frank with parameter $\theta=8$]{\includegraphics[height=5cm, width=5cm]{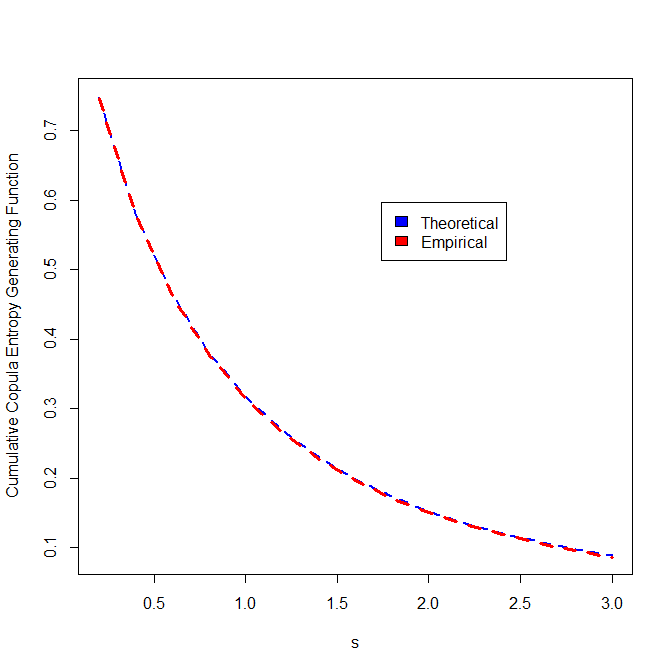}} \hfill
    \subfloat[Joe copula with parameter $\theta=1.5$]{\includegraphics[height=5cm, width=5cm]{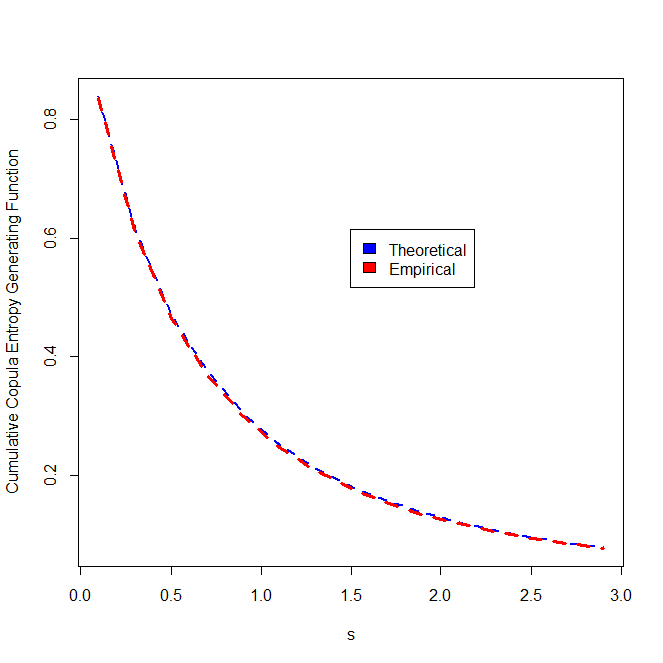}} 
    \caption{The empirical beta  and theoretical CCIGF of various bivariate copulas.}\label{fig3}
\end{figure}

\begin{figure}[H]
    \centering
      \subfloat[Product Copula]{\includegraphics[height=5cm, width=5cm]{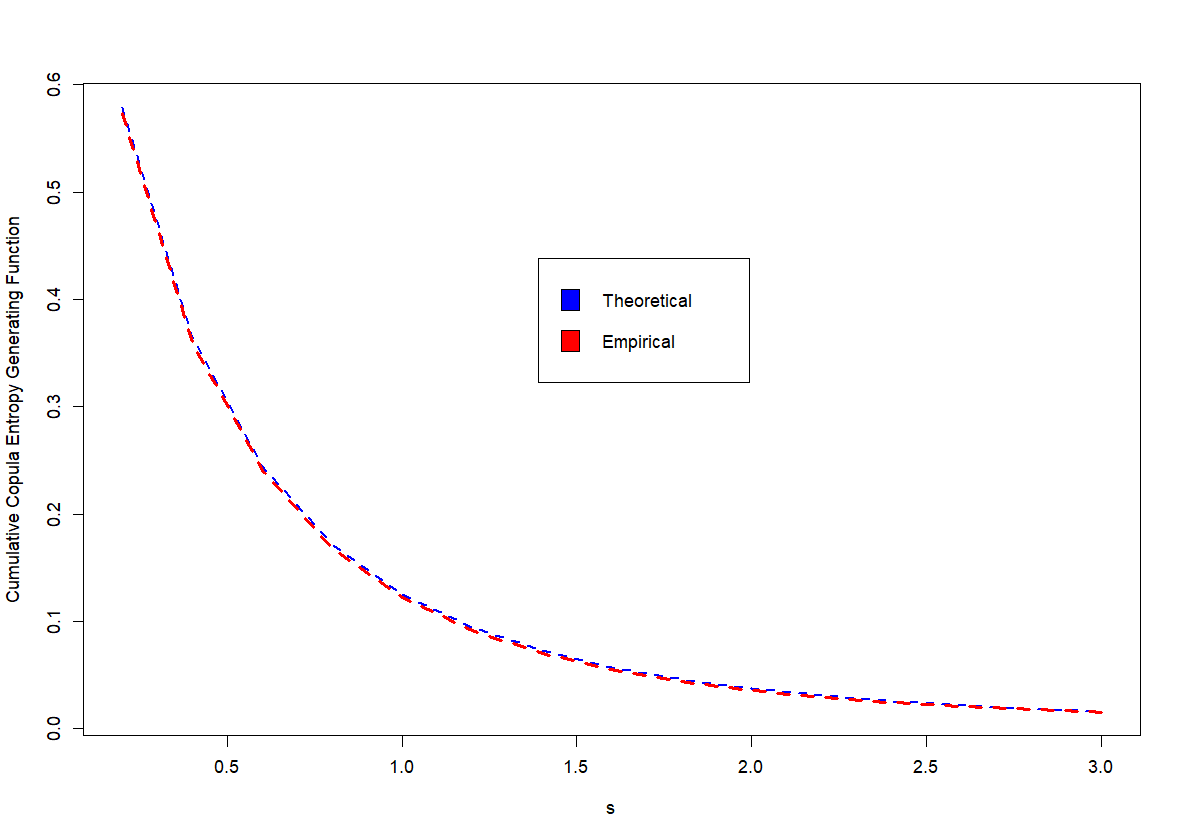}}\hfill
    \subfloat[Frank with parameter $\theta=8$]{\includegraphics[height=5cm, width=5cm]{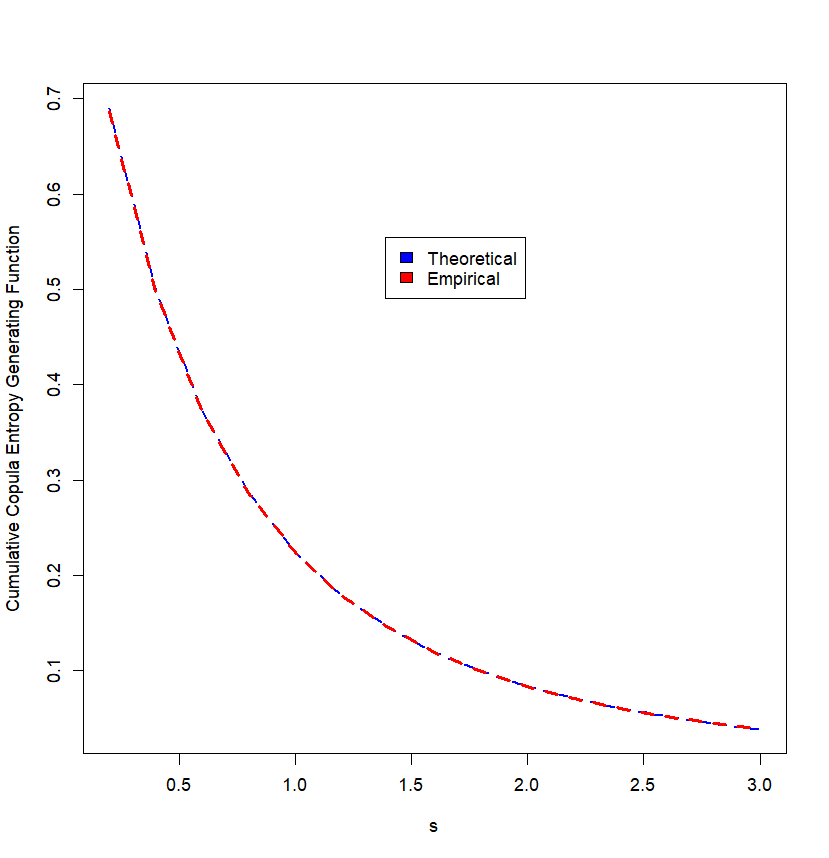}} \hfill
    \subfloat[Joe copula with parameter $\theta=1.5$]{\includegraphics[height=5cm, width=5cm]{cce/joedim2,1.5.png}} 
  
    \caption{The empirical beta  and theoretical CCIGF of various trivariate copulas.}\label{fig4}
\end{figure}

 We generated $1000$ random samples from the copula and computed the fractional empirical beta CCE and empirical beta generating function, comparing them with the actual values. Since no closed-form expression can be obtained for the empirical CCE, we evaluate the integrals numerically using the \texttt{adaptIntegrate} function in the \texttt{cubature} package of R (version 4.2.2).
The Figure \ref{fig1} and Figure \ref{fig2} shows the non-parametric estimate of fractional CCE for the Clayton copula, Gumbel-Hougaard copula, and Gaussian copula for bivariate and trivariate cases. From Figure \ref{fig1} and Figure \ref{fig2}, we can see that the shape of the fractional CCE varies with the dimension of the copula. The Figure \ref{fig3} and Figure \ref{fig4} show the consistency of the empirical  and its theoretical values for both bivariate and trivariate cases for various values of $s$. 
The following theorem will provide bound for the empirical .


\newpage
\begin{theorem}
     For any $k$-dimensional copula $C$, 
     %following inequality holds.
  % \begin{enumerate}
  % \item  $\zeta_r(\hat{C}_N) \leq \min\left\{\left(\dfrac{r}{e}\right)^{r}, \mathcal{A}_r(k)\right\}$ holds, where $\mathcal{A}_r(k) = \dfrac{\Gamma(r+1)}{(k-1)!}\left(\displaystyle\sum_{i=0}^{k-1}\binom{k-1}{i}\frac{(-1)^i}{(i+1)^{r+1}}\right)$ and $r\in\mathbb{I}$.

   $$\mathcal{G}_{\hat{C}_N}(s)  \begin{cases} 
\geq \left[\displaystyle N^{-1}\sum_{i=1}^N\prod_{j=1}^d\sum_{k=R_{i,j}}^N\beta(k+1,N-k+1)\right]^s, & \text{if } \  s > 1 \\
\leq \left[\displaystyle N^{-1}\sum_{i=1}^N\prod_{j=1}^d\sum_{y=R_{i,j}}^N\beta(y+1,N-y+1)\right]^s, & \text{if} \hspace{0.2cm} 0\leq s\leq 1.
\end{cases}$$
 %  \end{enumerate}
\end{theorem}
\begin{proof}
    %Since $\hat{C}_N(\mathbf{u})$ is a valid copula the proof of the first part of this theorem follows from theorem \ref{prop3.2}. Now 
    Consider the integral, 
    \begin{align*}
        \int_{\mathbb{I}^k}\hat{C}_N(\mathbf{u})d\mathbf{u}=&\int_{0}^{1}\int_{0}^{1} \cdots\int_{0}^{1}\dfrac{1}{N}\sum_{i=1}^N\prod_{j=1}^{k}\sum_{y=R_{i,j}}^N\binom{N}{y}u_j^y(1-u_j)^{N-y}du_1du_2\dots u_k\\
        =&\sum_{i=1}^N\prod_{j=1}^{k}\sum_{y=R_{i,j}}^N\beta(y+1,N-y+1).
    \end{align*}
    The proof now follows from Theorem \ref{prop4.1}.
\end{proof}
\section{Cumulative Copula Kullback-Leibler Divergence and its Application} \label{sec7}
In this section, we propose a new divergence measure between copulas based on Kullback-Leibler divergence proposed by \cite{kullback1951information}. Kullback and Leibler \cite{kullback1951information} proposed a discrimination measure between two random variables $X$ and $Y$, having PDF $f(x)$ and $g(x)$ respectively, defined as
\begin{equation}\label{dvg}
    D(f:g) = \int_{-\infty}^{\infty} f(x)\ln\left(\frac{f(x)}{g(x)}\right)dx.
\end{equation} This measure can be extended to higher dimension also.
Let $D_1(\cdot)$ and $D_2(\cdot)$ be the underlying copula density corresponds to random vectors $\mathbf{X}$ and $\mathbf{Y}$, respectively. Assume that the each component of $\mathbf{X}$ and $\mathbf{Y}$ are identically distributed, then using Eq. (\ref{pdf}) Kullback-Leibler divergence between two random vectors is 
\begin{equation}
    D(f:g) = \int_{\mathbb{R}^k} f(\mathbf{x})\ln\left(\frac{f(\mathbf{x})}{g(\mathbf{x})}\right)d\mathbf{x}=\int_{\mathbb{I}^k} D_1(\mathbf{u})\ln\left(\frac{D_1(\mathbf{u})}{D_2(\mathbf{u})}\right)d\mathbf{u}.
\end{equation}
Thus, Kullback-Leibler divergence between two random vectors can be expressed in terms of copula density under certain conditions. Main limitation of this divergence measure is the existence of copula density. In many situations copula density need not exist, this motivated us to propose a new divergence measure in terms of cumulative copula which measure the divergence between two copulas. Baratpour and Rad \cite{baratpour2012testing} proposed a new distance measure based on the survival function of two non-negative random variables. Let $\bar{F}(x)$ and $\bar{G}(x)$ be the survival functions of $X$ and $Y$, respectively. Then, cumulative Kullback-Leibler (CKL) divergence of $X$ and $Y$ is given by
 \[CKL(F:G) = \int_{0}^{\infty} \bar{F}(x)\ln\left(\frac{\bar{F}(x)}{\bar{G}(x)}\right)dx-\left[E(X)-E(Y)\right].\] 
Baratpour and Rad \cite{baratpour2012testing} used this measure for the goodness of fit test for exponential distribution. Inspired by the work of \cite{baratpour2012testing}, we are propose a new distance measure between two copulas of the same dimension.
 \begin{definition}
     Let $C_1(\mathbf{u})$ and $C_2(\mathbf{u})$ be two copulas of same dimension, then the cumulative copula Kullback-Leibler (CCKL) divergence of $C_1(\mathbf{u})$ and $C_2(\mathbf{u})$ is defined as
     \begin{equation}\label{cckl}
         CCKL(C_1:C_2)=\int_{\mathbb{I}^k} C_1(\mathbf{u})\ln\left(\frac{C_1(\mathbf{u})}{C_2(\mathbf{u})}\right)d\mathbf{u}-\left[\dfrac{\rho_k^{-}(C_1)-\rho_k^{-}(C_2)}{2^kn(k)}\right].
     \end{equation}
 \end{definition}
 The following theorem confirms the proposed CCKL divergence, a well-defined distance measure, between two copulas. 
 \begin{theorem}\label{pr2}
     $CCKL(C_1:C_2)\geq0$ and equality holds if and only if $C_1(\mathbf{u})=C_2(\mathbf{u}), \ \forall \mathbf{u}\in\mathbb{I}^k $.
 \end{theorem}
 \begin{proof}
     Using the inequality $x\ln\left(\dfrac{x}{y}\right)\geq x-y$ for every non-negative $x$ and $y$ and by definition of multivariate version of Spearman's Rho, we have
     \begin{align*}
        \int_{\mathbb{I}^k} C_1(\mathbf{u})\ln\left(\frac{C_1(\mathbf{u})}{C_2(\mathbf{u})}\right)d\mathbf{u}-\left[\dfrac{\rho_k^{-}(C_1)-\rho_k^{-}(C_2)}{2^kn(k)}\right] \geq\int_{\mathbb{I}^k} C_1(\mathbf{u})-C_2(\mathbf{u})d\mathbf{u}-\left[\dfrac{\rho_k^{-}(C_1)-\rho_k^{-}(C_2)}{2^kn(k)}\right] \geq 0.
     \end{align*}
     It is straight forward that if $C_1(\mathbf{u})=C_2(\mathbf{u})$ then $CCKL(C_1:C_2)=0$. Conversely, suppose that $CCKL(C_1:C_2)=0$, it follows that 
     \begin{align*}
     0&=\int_{\mathbb{I}^k} C_1(\mathbf{u})\ln\left(\frac{C_1(\mathbf{u})}{C_2(\mathbf{u})}\right)d\mathbf{u}-\left[\dfrac{\rho_k^{-}(C_1)-\rho_k^{-}(C_2)}{2^kn(k)}\right]\\
        &=\int_{\mathbb{I}^k} C_1(\mathbf{u})\ln\left(\frac{C_1(\mathbf{u})}{C_2(\mathbf{u})}\right)-\left[C_1(\mathbf{u})-C_2(\mathbf{u})\right]d\mathbf{u} \\
        &=\int_{\mathbb{I}^k} \left[\dfrac{C_2(\mathbf{u})}{C_1(\mathbf{u})}-\ln\left(\dfrac{C_2(\mathbf{u})}{C_1(\mathbf{u})}\right)-1\right]C_1(\mathbf{u})d\mathbf{u}.
     \end{align*} 
     It is easy to verify that $g(z)=z-\ln(z)-1$ is non-negative for every $z\geq0$ and $g(z)=0$ if and only if $z=1$. It follows that $C_1(\mathbf{u})=C_2(\mathbf{u}),$ for every $\mathbf{u}\in\mathbb{I}^k $.
 \end{proof}
 Now, we will consider the CCKL divergence of some well known copulas.
 \begin{example}
     Consider the Fr{\'e}chet-Hoeffding lower bound copula $W(u_1,u_2)=\max\{u_1+u_2-1,0\}$ and the product copula $\Pi(u_1,u_2)=u_1u_2$. It is obvious that $\rho_2^{-}(W)=-1$ and $\rho_2^{-}(\Pi)=0$. Moreover,
     \begin{align*}
         \int_0^1\int_0^1 W(u_1,u_2)\ln\left(\dfrac{W(u_1,u_2)}{u_1u_2}\right)du_1du_2=&\int_0^1\int_{1-u_1}^1 (u_1+u_2-1)\ln\left(\dfrac{u_1+u_2-1}{u_1u_2}\right)du_2du_1
         =\dfrac{1}{36},
     \end{align*} 
    and $n(2)=3$ (using Eq. \ref{cckl}). It follows that the CCKL between the Fr{\'e}chet-Hoeffding lower bound copula and product copula is $\dfrac{1}{36}+\dfrac{1}{12}=\dfrac{1}{9}.$
 \end{example}
 \begin{example}
     Consider the bivariate product copula $\Pi(u_1,u_2)=u_1u_2$ and the Gumbel-Barnett copula
     $$C(u_1,u_2)=u_1u_2\exp\{-\theta\ln(u_1)\ln(u_2)\}, \theta\in\mathbb{I}.$$ Since $\rho_2^{-}(C)=-12\left(\theta^{-1}e^{4/\theta}E_i\left(-4/\theta\right)\right)-3$ (see \cite{yela2018estimating}), where $E_i(\cdot)$ is the usual exponential integral function. Consider the integral
     \begin{align*}
         \int_0^1\int_0^1 u_1u_2\ln\left(\dfrac{u_1u_2}{u_1u_2\exp\{-\theta\ln(u_1)\ln(u_2)\}}\right)du_1du_2=\dfrac{\theta}{16}.
     \end{align*}
     Therefore, the CCKL between the product copula and Gumbel-Barnett copula is given by
     \begin{align*}
         CCKL(\Pi:C)=\dfrac{\theta}{16}-\dfrac{12\left(\theta^{-1}e^{4/\theta}E_i\left(-4/\theta\right)\right)-3}{2^2 n(2)}=\dfrac{\theta}{16}-\left(\theta^{-1}e^{4/\theta}E_i\left(-4/\theta\right)+\dfrac{1}{4}\right),\theta\in\mathbb{I}.
     \end{align*}
 \end{example}
 \begin{example}
    Consider the $k-$dimensional product copula $\Pi(\mathbf{u})=u_1u_2\dots u_k$ and the minimum copula $M(\mathbf{u})=\min\{u_1,u_2,\dots,u_k\}$. We have $\rho_k^{-}(\Pi)=0$ and $\rho_k^{-}(M)$ is 
    \begin{align*} 
 \rho_k^{-}(M)=&n(k)\left[2^k\int_{\mathbb{I}^k}M(\mathbf{u})d\mathbf{u}-1\right]\\
=&n(k)\left[2^k\int_{0}^1ku(1-u)^{k-1}du-1\right] \hspace{0.5cm} \text{(using Eq. (\ref{order}))}\\
=&n(k)\left[2^k\beta(2,k)-1\right],
\end{align*}
where $\beta(p,q)$ is the usual beta function. Moreover,
\begin{align*}
\int_{\mathbb{I}^k} \Pi(\mathbf{u})\ln\left(\frac{\Pi(\mathbf{u})}{M(\mathbf{u})}\right)d\mathbf{u}=&\int_0^1\int_0^1\cdots\int_0^1u_1u_2\dots u_k\ln\left(\frac{u_1u_2\dots u_k}{\min\{u_1,u_2,\dots,u_k\}}\right)du_1du_2\dots du_k\\
=&k!\int_0^1\int_0^{u_1}\int_0^{u_2}\cdots\int_0^{u_{k-1}}u_1u_2\dots u_k\ln\left(u_2u_3\dots u_k\right)du_1du_2\dots du_k\\
=&\sum_{i=2}^{k}J_i,
 \end{align*}
where for each $i =2,3,\cdots, k$, $J_i=k!\int_0^1\int_0^{u_k}\int_0^{u_{k-1}}\cdots\int_0^{u_{2}}u_1u_2\dots u_k\ln\left(u_i\right)du_1du_2\dots du_k=-2^{-k-1}\sum_{n=i}^kn^{-1}$. Then the CCKL divergence between the product copula and minimum copula is given by
\begin{align*}
     CCKL(\Pi:M)=&\int_{\mathbb{I}^k} \Pi(\mathbf{u})\ln\left(\frac{\Pi(\mathbf{u})}{M(\mathbf{u})}\right)d\mathbf{u}-\left[\dfrac{\rho_k^{-}(\Pi)-\rho_k^{-}(M)}{2^kn(k)}\right]. \\
     =&\sum_{i=2}^{k}J_i+\beta(2,k)-2^{-k}.
    \end{align*}
 \end{example}
 \begin{example}
     Consider the $k$-dimensional Cuadras-Aug{\'e} copula defined in Eq. (\ref{cu}) and the minimum copula. The multivariate Spearman's Rho of Cuadras-Aug{\'e} copula is given by
     \begin{align*}
      \rho_k^{-}(C)=& n(k)\left[2^k\int_{\mathbb{I}^k}\prod_{i=1}^ku_{(i)}^{\theta_i}d\mathbf{u}-1\right]\\
      =&n(k)\left[\dfrac{2^k}{\prod_i^kp(i)}-1\right], 
     \end{align*}
     where $u_{(1)}\leq u_{(2)}\cdots\leq u_{(k)}$, $\theta_1 = 1$, $\theta_i = \prod_{j=1}^{i-1}(1 - \alpha_{ij})$, and $p(i) = p(i-1) + \theta_i + 1$ with $p(1) = 2$, for $i = 2, 3, \dots, k$. Furthermore, 
     \begin{align*}
         \int_{\mathbb{I}^k} C(\mathbf{u})\ln\left(\frac{C(\mathbf{u})}{M(\mathbf{u})}\right)d\mathbf{u}=&\int_0^1\int_0^1\cdots\int_0^1u_{(1)}u_{(2)}^{\theta_2}\dots u_{(k)}^{\theta_k}\ln\left(\frac{u_{(1)}u_{(2)}^{\theta_2}\dots u_{(k)}^{\theta_k}}{u_{(1)}}\right)\ du_1du_2\dots du_k\\
=&k!\sum_{j=2}^{k}\int_0^1\int_0^{u_k}\int_0^{u_{k-1}}\cdots\int_0^{u_{2}}u_1u_2^{\theta_2}\dots u_k^{\theta_k}\ln\left(u_j^{\theta_j}\right)du_1du_2\dots du_k\\
=&-k!\sum_{j=2}^{k}\theta_jI_j,
     \end{align*}
     where $I_j=\dfrac{1}{\prod_{i=1}^kp(i)}\left(\displaystyle\sum_{i=j}^k\dfrac{1}{p(j)}\right)$ for every $j=2,3,\dots,k$. The CCKL divergence between the Cuadras-Aug{\'e} copula and minimum copula is given by 
     $$CCKL(C:M)=-k!\sum_{j=2}^{k}\theta_jI_j-\dfrac{1}{\prod_i^kp(i)}+\beta(2,k).$$
 \end{example}
In the literature, several bootstrapping test procedures exist for the goodness-of-fit test for copulas (see \cite{GF2,gF1,gf3}). In the following subsection, we propose a goodness-of-fit test procedure for copulas based on the cumulative copula Kullback-Leibler distance as an application.
\subsection{A Goodness of fit test for copula}
Let $\{C_{\theta}:\theta\in\Theta\}$ be a family of copula functions. we want to test the hypothesis 
$$H_0 \ : \ C=C_{\theta}, \quad vs \quad H_A \ : \ C\neq C_{\theta}.$$  Now, 
Using the definition of CCKL divergence between two copulas, the above hypothesis is equivalent to the hypothesis
$$H_0 \ : \ CCKL(C:C_{\theta})=0, \quad vs \quad H_A \ : \ CCKL(C:C_{\theta})>0.$$
The CCKL distance between $C$ and $C_{\theta}$ is given by 
 \begin{align*}
    CCKL(C:C_{\theta})=&\int_{\mathbb{I}^k} C(\mathbf{u})\ln\left(\frac{C(\mathbf{u})}{C_{\theta}(\mathbf{u})}\right)d\mathbf{u}-\left[\dfrac{\rho_k^{-}(C)-\rho_k^{-}(C_{\theta})}{2^kn(k)}\right]\\
    =&\int_{\mathbb{I}^k} C(\mathbf{u})\ln\left(\frac{C(\mathbf{u})}{C_{\theta}(\mathbf{u})}\right)- C(\mathbf{u})+C_{\theta}(\mathbf{u}) \ d\mathbf{u}\\
    =& -\zeta(C)- \int_{\mathbb{I}^k} C(\mathbf{u})\ln\left(C_{\theta}(\mathbf{u})\right)d\mathbf{u}-\int_{\mathbb{I}^k} C(\mathbf{u})d\mathbf{u}+\int_{\mathbb{I}^k} C_{\theta}(\mathbf{u})d\mathbf{u}. 
 \end{align*}
 Let $\mathbf{X}_1,\mathbf{X}_2,\dots, \mathbf{X}_N$ be $N$ random samples from a $k$-variate distribution with underlying copula $C$. We approximate the copula $C$ by empirical beta copula $\hat{C}_N$ which yields the following test statistic 
 \begin{equation}\label{teststat1}
    T_N = -\zeta(\hat{C}_N)- \int_{\mathbb{I}^k} \hat{C}_N(\mathbf{u})\ln\left(C_{\theta}(\mathbf{u})\right)d\mathbf{u}-\int_{\mathbb{I}^k} \hat{C}_N(\mathbf{u})d\mathbf{u}+\int_{\mathbb{I}^k} C_{\theta}(\mathbf{u})d\mathbf{u}.
\end{equation}
\par Using theorem \ref{pr1}, we showed that $\zeta(\hat{C}_N)$ is a consistent estimator for $\zeta(C_{\theta})$ under $H_0$. It implies that 
$T_N \xrightarrow{P} 0$ under $H_0$. Moreover, since $\hat{C}_N$ is also a valid copula and by theorem \ref{pr2}, we have $CCKL(\hat{C}_N:C)>0,$ under $H_A$. Consequently, $P(T_N> 0)=1$ as $N\rightarrow \infty$ under $H_A$. Therefore, the test based on the test statistic  $T_N$ is a consistent test. We reject the null hypothesis $H_0$ at  significance level $\alpha$ if $T_N\geq T_{N,1-\alpha}$, where $T_{N,1-\alpha}$ is the $100(1-\alpha)$th percentile of $T_N$ under $H_0$. The distribution of $T_N$ under $H_0$ can't be obtained analytically, so the Monte Carlo simulation method will be used to determine the the value of $T_{N,1-\alpha}$. Since computing the test statistic \( T_N \) in Eq. (\ref{teststat1}) is often time-consuming, especially for large values of \( N \), we need to approximate \( T_N \) by its sample counterpart. Note that the right-hand side (RHS) of Eq. (\ref{teststat1}) can be expressed as 
\[
E\left[C_N(\mathbf{U}) \ln \left(C_N(\mathbf{U}) \right) - C_N(\mathbf{U}) \ln \left(C_{\theta}(\mathbf{U}) \right) - C_N(\mathbf{U}) + C_{\theta}(\mathbf{U}) \right],
\]
where \( U_1, U_2, \dots, U_k \) are \( k \) independently and identically distributed random variables from a uniform distribution over \( \mathbb{I} \) and \( \mathbf{U} = (U_1, U_2, \dots, U_k) \). We approximate the expectation by the sample mean, which yields the approximate value of the test statistic \( T_N \) given by
\begin{equation}\label{teststat}
    T_N = \frac{1}{N} \sum_{i=1}^{N} \left[ \hat{C}_N(\mathbf{e}_i) \ln \left( \hat{C}_N(\mathbf{e}_i) \right) - \hat{C}_N(\mathbf{e}_i) \ln \left( \hat{C}_{\theta}(\mathbf{e}_i) \right) - \hat{C}_N(\mathbf{e}_i) + C_{\theta}(\mathbf{e}_i) \right],
\end{equation}
where \( \mathbf{e}_i = (e_{i,1}, e_{i,2}, \dots, e_{i,k}) \), with \( e_{i,j} = \frac{R_{i,j}}{N+1} \), and \( R_{i,j} \) is the rank of the \( i \)-th observation of the \( j \)-th component \( X_{i,j} \) for \( i = 1, 2, \dots, N \) and \( j = 1, 2, \dots, k \). The following algorithm will provide an estimated p-value and $100(1-\alpha)$th percentile of $T_N$ for the proposed test based on the statistic $T_N$. 
\begin{enumerate}[Step 1:]
\item Estimate the copula parameter $\theta$ from the given data of size $N$. We can use any consistent estimator $\hat{\theta}_N$ of copula parameter $\theta$.
	\item Compute the value of the test statistic $T_N$ given in the Eq. (\ref{teststat}).
	\item Generate $M$ random samples of size $N$ from the copula with copula parameter $\hat{\theta}_N$, estimate $\theta$ by the same consistent estimator used in Step 1, and calculate the test statistic for each random sample.
 \item Let $T_{N_{(1)}},T_{N_{(2)}},\dots,T_{N_{(M)}}$ be the ordered values of the computated test statistic $T_N$ in Step 3. Then the estimated $100(1-\alpha)$th percentile value of $T_N$ is
    $T_{N_{\left[(1-\alpha)M)\right]}}$, where $[\cdot]$ denotes greatest integer function.
    \item The estimated $p$-value associated with the observed test statistic $T_N$ can be computed by  $$\displaystyle\frac{1}{M}\sum_{j=1}^{M}\mathbf{1}\left\{r:T_{N_{(r)}}\geq T_N\right\}.$$
   
 \end{enumerate}
\section{Simulation Study and Data Analysis} \label{sec6}
In this section, an extensive simulation study is conducted to estimate the $95$th percentile of the test statistic \(T_N\) for various sample sizes under different copula models.
%Additionally, we discuss the power of the proposed test for these copula models.
The simulation study is performed using R software (version 4.2.2). For simulation study we generated $10,000$ samples of size $N=100,150,200$ and $250$ from different copulas.
\par First, we estimate the $95$th percentile of $T_N$ based on sample sizes $N=100,150,200$ and $250$ under Clayton, Frank, Gumbel-Hougaarad, Joe, Normal, and product copulas. The estimated 95th percentile of the statistic \(T_N\) for the bivariate and trivariate versions of the considered copulas are reported in Table \ref{t1} and Table 
 \ref{t2}, respectively. 

\begin{table}[ht]
\centering
\caption{Estimated values of the $95$th percentile of \(T_N\) for various bivariate copula models}\label{t1}
\footnotesize
\begin{tabular}{llllll}
\hline
\multirow{2}{*}{Model}   & \multirow{2}{*}{Parameter} & \multicolumn{4}{c}{Sample Size} \\
                         &                            & 100                     & 150                     & 200                     & 250                     \\ \hline
\multirow{1}{*}{Clayton} & $\alpha = 0.5$             & $1.4062 \times 10^{-3}$ & $9.9775 \times 10^{-4}$ & $7.4645 \times 10^{-4}$ & $5.9598 \times 10^{-4}$ \\
                         & $\alpha = 2$               & $7.5188 \times 10^{-4}$ & $4.8142 \times 10^{-4}$ & $3.5280 \times 10^{-4}$ & $2.8429 \times 10^{-4}$ \\
                         & $\alpha = 6$               & $3.8466 \times 10^{-4}$ & $2.2941 \times 10^{-4}$ & $1.6636 \times 10^{-4}$ & $1.2723 \times 10^{-4}$ \\
\multirow{1}{*}{Frank}   & $\theta = 3$               & $1.3110 \times 10^{-3}$ & $8.9599 \times 10^{-4}$ & $7.1967 \times 10^{-4}$ & $5.7994 \times 10^{-4}$ \\
                         & $\theta = 5$               & $1.05197 \times 10^{-3}$& $7.0819 \times 10^{-4}$ & $5.2954 \times 10^{-4}$ & $4.3991 \times 10^{-4}$ \\
                         & $\theta = 14$              & $4.9869 \times 10^{-4}$ & $3.3738 \times 10^{-4}$ & $2.5221 \times 10^{-4}$ & $2.0387 \times 10^{-4}$ \\
\multirow{1}{*}{Gumbel-Hougaarad}  & $\phi = 1.5$               & $1.3802 \times 10^{-3}$ & $9.6353 \times 10^{-4}$ & $7.3333 \times 10^{-4}$ & $5.8284 \times 10^{-4}$ \\
                         & $\phi = 2$                 & $1.0113 \times 10^{-3}$ & $7.0783 \times 10^{-4}$ & $5.3709 \times 10^{-4}$ & $4.2715 \times 10^{-4}$ \\
                         & $\phi = 4$                 & $5.0686 \times 10^{-4}$ & $3.2607 \times 10^{-4}$ & $2.3746 \times 10^{-4}$ & $1.1916 \times 10^{-4}$ \\
\multirow{1}{*}{Joe}     & $\theta = 1.5$             & $1.6839 \times 10^{-3}$ & $1.1841 \times 10^{-3}$ & $9.2153 \times 10^{-4}$ & $7.4284 \times 10^{-4}$ \\
                         & $\theta = 3$               & $1.1915 \times 10^{-3}$ & $8.5074 \times 10^{-4}$ & $6.4118 \times 10^{-4}$ & $5.2301 \times 10^{-4}$ \\
                         & $\theta = 7$               & $7.0373 \times 10^{-4}$ & $4.8555 \times 10^{-4}$ & $3.6759 \times 10^{-4}$ & $3.0002 \times 10^{-4}$ \\
\multirow{1}{*}{Normal}  & $\rho = 0.4$               & $1.4116 \times 10^{-3}$ & $9.3985 \times 10^{-4}$ & $7.2460 \times 10^{-4}$ & $5.7744 \times 10^{-4}$ \\
                         & $\rho = 0.7$               & $8.7301 \times 10^{-4}$ & $5.9016 \times 10^{-4}$ & $4.4601 \times 10^{-4}$ & $3.5730 \times 10^{-4}$ \\
                         & $\rho = 0.9$               & $4.6217 \times 10^{-4}$ & $3.0072 \times 10^{-4}$ & $2.2217 \times 10^{-4}$ & $1.7607 \times 10^{-4}$ \\
Product                  &                            & $2.0239 \times 10^{-3}$ & $1.3582 \times 10^{-3}$ & $1.0457 \times 10^{-3}$ & $8.3575 \times 10^{-4}$ \\ \hline
\end{tabular}
\end{table}
% \usepackage{multirow}
% \usepackage[normalem]{ulem}
\begin{table}[ht]
\centering
\caption{Estimated values of the $95$th percentile of \(T_N\) for various trivariate copula models}\label{t2}
\footnotesize
\begin{tabular}{llllll}
\hline
\multirow{2}{*}{Model}   & \multirow{2}{*}{Parameter} & \multicolumn{4}{c}{Sample Size} \\
                         &                            & 100                     & 150                     & 200                     & 250                     \\ \hline
\multirow{1}{*}{Clayton} 
    & $\alpha=0.5$ & $2.7319\times 10^{-3}$ & $1.8752\times 10^{-3}$ & $1.3945\times 10^{-3}$ & $1.11761\times 10^{-3}$ \\
    & $\alpha=2$   & $1.5505\times 10^{-3}$ & $9.9921\times 10^{-4}$ & $7.2811\times 10^{-4}$ & $5.5819\times 10^{-4}$ \\
    & $\alpha=6$   & $8.3354\times 10^{-4}$ & $4.8876\times 10^{-4}$ & $3.4780\times 10^{-4}$ & $2.6978\times 10^{-4}$ \\
\multirow{1}{*}{Frank} 
    & $\theta=3$   & $2.3452\times 10^{-3}$ & $1.6175\times 10^{-3}$ & $1.2467\times 10^{-3}$ & $9.9976\times 10^{-4}$ \\
    & $\theta=5$   & $1.9130\times 10^{-3}$ & $1.3126\times 10^{-3}$ & $9.7085\times 10^{-4}$ & $8.0744\times 10^{-4}$ \\
    & $\theta=14$  & $1.0311\times 10^{-3}$ & $6.7041\times 10^{-4}$ & $5.0303\times 10^{-4}$ & $4.0423\times 10^{-4}$ \\
\multirow{1}{*}{Gumbel-Hougaarad} 
    & $\phi=1.5$   & $2.5235\times 10^{-3}$ & $1.7071\times 10^{-3}$ & $1.2952\times 10^{-3}$ & $1.0597\times 10^{-3}$ \\
    & $\phi=2$     & $1.9748\times 10^{-3}$ & $1.31551\times 10^{-3}$ & $1.0025\times 10^{-3}$ & $8.0576\times 10^{-4}$ \\
    & $\phi=4$     & $1.0649\times 10^{-3}$ & $6.8730\times 10^{-4}$ & $5.0277\times 10^{-4}$ & $3.898\times 10^{-4}$ \\
\multirow{1}{*}{Joe} 
    & $\theta=1.5$ & $2.9452\times 10^{-3}$ & $2.0107\times 10^{-3}$ & $1.5395\times 10^{-3}$ & $1.2463\times 10^{-3}$ \\
    & $\theta=3$   & $2.23767\times 10^{-3}$ & $1.5386\times 10^{-3}$ & $1.1609\times 10^{-3}$ & $9.3941\times 10^{-4}$ \\
    & $\theta=7$   & $1.4641\times 10^{-3}$ & $9.5634\times 10^{-4}$ & $7.3481\times 10^{-4}$ & $5.9205\times 10^{-4}$ \\
\multirow{1}{*}{Normal} 
    & $\rho=(0.1,0.2,0.3)$ & $2.7463\times 10^{-3}$ & $1.9346\times 10^{-3}$ & $1.4515\times 10^{-3}$ & $1.1740\times 10^{-3}$ \\
    & $\rho=(0.4,0.5,0.6)$ & $2.1174\times 10^{-3}$ & $1.4448\times 10^{-3}$ & $1.0955\times 10^{-3}$ & $9.0260\times 10^{-4}$ \\
    & $\rho=(0.7, 0.8, 0.9)$ & $1.2762\times 10^{-3}$ & $8.1735\times 10^{-4}$ & $6.2295\times 10^{-4}$ & $4.6812\times 10^{-4}$ \\
Product & & $3.066207\times 10^{-3}$ & $2.1759\times 10^{-3}$ & $1.6934\times 10^{-3}$ & $1.3827\times 10^{-3}$ \\ \hline
\end{tabular}
\end{table}
The power and size of the proposed test were estimated using $10,000$ simulations across various sample sizes and copula models, and the results are presented in the appendix.
%Next, we will focus on the power and size of the proposed test for different copulas.
It is observed that as the dimension of the copula increases, power of the proposed test also increased in most cases. In order to continue our discussion, in the following subsection we use the proposed test for the copula selection problem to a real data set.
\subsection{Selection of an appropriate copula for ``Pima Indians Diabetes" data}
In this subsection, we analyze a real dataset to demonstrate the practical utility of the copula selection problem. We consider the ``Pima Indians Diabetes" data. The US National Institute of Diabetes and Digestive and Kidney Diseases collected diabetes data from women aged 21 and above, who were of Pima Indian descent and lived around Phoenix, Arizona. The data is freely available in the R software within the \texttt{pdp} package. We consider the variables ``glucose", ``pressure", and ``mass" from the dataset, which represent plasma glucose concentration, diastolic blood pressure (mm Hg), and body mass index, respectively. All missing values were removed, resulting a trivariate dataset with $724$ entries. The copulas considered in this study include Clayton, Frank, Gumbel-Hougaarad, Joe, Normal and product copula. The marginal CDF are estimated by empirical distribution and copula parameters are estimated using maximum psuedo-likelihood (mpl) estimation method. We use our proposed method for the goodness of fit test and the p-values of the proposed test are estimated for each copula models. We use the copula having least CCKL distance defined in Eq. (\ref{cckl}) between the empirical beta copula is considered as the model selection criteria. We generated $1000$ random samples of size $N=724$ for estimating the p-values. The MPL estimates of the copula parameters, CCKL value and p-values are reported in Table \ref{t9}. From Table \ref{t9}, Frank copula have the least CCKL distance between empirical beta copula with p-value$=0.448$. It follows that Frank copula can be considered as an appropriate choice for modelling the given dataset. 
\begin{table}[H]
\centering
\caption{MPL Estimates of copula, CCKL distance and p-values of the proposed test}\label{t9}
\begin{tabular}{llll}
\hline
Copula           & Estimate                  & CCKL                   & p-value \\ \hline

Clayton          & $0.23907$                 & $5.9516\times10^{-4}$   & 0.029   \\
Frank            & $1.37762$                 & $2.4595\times10^{-4}$  & 0.488   \\
Gumbel-Hougaarad & $1.15423$                 & $5.2801\times 10^{-4}$ & 0.036   \\
Joe              & $1.19773$                 & $1.4895\times10^{-3}$  & 0       \\
Normal           & $0.22994,0.22644,0.28999$ & $3.7083\times 10^{-4}$ & 0.13    \\
Product          &                           & $1.4921\times10^{-3}$  & 0       \\ \hline
\end{tabular}
\end{table}
\section{Conclusion} \label{cocl}
In this paper, we discuss the multivariate version of the CCE and study its various mathematical properties. We propose the CCIGF and explore its properties. Using fractional calculus, we
introduce a fractional version of the multivariate cumulative copula entropy. We proved that the CCE of the weighted arithmetic mean of copulas never exceeds the weighted arithmetic mean of the CCE of copulas. The results are valid for the CCIGF and FCCE. We showed that concordance ordering for two copulas never implies entropy ordering by a counter-example and provides conditions for the entropy ordering of two copulas. The results are valid for FCCE. However, in the case of CCIGF, concordance ordering preserves the ordering of corresponding CCIGF. We also showed that the CCIGF of the weighted geometric mean of copulas never exceeds the weighted geometric mean of the CCIGF of copulas. We provide a non-parametric estimate of the FCCE and CCIGF using the empirical beta copula. We showed that the proposed non-parametric estimate converges almost surely to the true FCCE and CCIGF, theoretically and numerically. We define a new distance measure between two copulas using the
Kullback-Leibler divergence. Furthermore, using the proposed distance measure, a goodness-of-fit test procedure is
proposed for copulas. A copula selection procedure is discussed
through the ``Pima Indians Diabetes'' dataset to illustrate the applications of the
new distance measure. 
%% Tices part is started with the command \appendix;


%% References

\section*{Declaration of interests}
The authors declare no potential conflict of interests.
\section*{Appendix}\label{append}
To calculate size and power, we generated $10,000$ samples of size $N=100,150,200$ and $250$ from the specific copula and estimated the size and power of the test based on whether or not the original data came from the assumed copula family under the null hypothesis. It is to be noted that in each bootstrapping sample, we assume that the copula parameters are known in advance, so we are not estimating copula parameter. Tables [\ref{t3}, \ref{t4}, \ref{t5}] shows the size and power of the test for some bivariate copula models and Tables [\ref{t6}, \ref{t7}, \ref{t8}] shows that size and power (in percentage) of the test for some trivariate copula models. Note that the size of the proposed test is given in bold format, and the copula model parameter values are mentioned in brackets next to each copula model.  
\begin{table}[H]
\centering

\caption{Percentage of rejection of $H_0$ for different bivariate copula models}\label{t3}
\scalebox{0.9}{
\begin{tabular}{llllll}
\hline
\multirow{2}{*}{Copula   under $H_0$} & \multirow{2}{*}{True Copula} & \multicolumn{4}{l}{Sample Size}                                   \\ \cline{3-6} 
                                      &                              & 100            & 150            & 200            & 250            \\ \hline
Clayton$(0.5)$                        & Clayton$(0.5)$               & \textbf{4.88}  & \textbf{5.14}  & \textbf{5.17}  & \textbf{4.81}  \\
                                      & Frank$(3)$                   & 26.57          & 48.59          & 65.42          & 79.61          \\
                                      & Gumbel-Hougaarad$(1.5)$                & 33.28          & 57.34          & 80.06          & 88.67          \\
                                      & Joe$(1.5)$                   & 57.9           & 80.54          & 91.82          & 97.15          \\
                                      & Normal$(0.4)$                & 11.2           & 16.73          & 23.33          & 32.4           \\
                                      & Product                      & 93.64          & 99.12          & 99.92          & 99.98          \\
                                      &                              &                &                &                &                \\
Frank$(3)$                            & Clayton$(0.5)$               & 41.21          & 61.87          & 75.95          & 85.6           \\
                                      & Frank$(3)$                   & \textbf{4.79}  & \textbf{5.13}  & \textbf{4.92}  & \textbf{5.27}  \\
                                      & Gumbel-Hougaarad$(1.5)$                & 7.29           & 8.62           & 8.83           & 10.43          \\
                                      & Joe$(1.5)$                   & 59.96          & 76.14          & 85.77          & 92.41          \\
                                      & Normal$(0.4)$                & 13.26          & 17.33          & 23.11          & 31.43          \\
                                      & Product                      & 99.4           & 99.97          & 100            & 100            \\
                                      &                              &                &                &                &                \\
Gumbel-Hougaarad$(1.5)$                         & Clayton$(0.5)$               & 41.33          & 64.49          & 80.4           & 88.88          \\
                                      & Frank$(3)$                   & 4.92           & 5.31           & 6.4            & 7.35           \\
                                      &Gumbel-Hougaarad$(1.5)$                & \textbf{5.97}  & \textbf{5.16}  & \textbf{4.94}  & \textbf{4.85}  \\
                                      & Joe$(1.5)$                   & 52.02          & 67.91          & 80.61          & 88.18          \\
                                      & Normal$(0.4)$                & 11.43          & 14.56          & 19.2           & 22.15          \\
                                      & Product                      & 99.33          & 99.97          & 100            & 100            \\
                                      &                              &                &                &                &                \\
Joe$(1.5)$                            & Clayton$(0.5)$               & 56.79          & 75.55          & 87.98          & 94.56          \\
                                      & Frank$(3)$                   & 44.82          & 65.32          & 80.01          & 97.79          \\
                                      & Gumbel-Hougaarad$(1.5)$                & 40.41          & 58.22          & 71.96          & 83.04          \\
                                      & Joe$(1.5)$                   & \textbf{5.23}           & \textbf{4.96}           & \textbf{4.83}           & \textbf{5.28}           \\
                                      & Normal$(0.4)$                & 28.41 & 44.37 & 57.38 & 69.64 \\
                                      & Product                      & 71.68          & 88.85          & 95.67          & 98.74          \\
                                      &                              &                &                &                &                \\
Normal$(0.4)$                         & Clayton$(0.5)$               & 14.3           & 21.6           & 30.74          & 36.78          \\
                                      & Frank$(3)$                   & 6.06           & 9.91           & 13.54          & 17.09          \\
                                      &Gumbel-Hougaarad$(1.5)$                & 8.01           & 12.14          & 15.39          & 19.71          \\
                                      & Joe$(1.5)$                   & 38.97          & 52.23          & 66.78          & 77.47          \\
                                      & Normal$(0.4)$                & \textbf{4.98}  & \textbf{5.24}  & \textbf{5.03}  & \textbf{4.88}  \\
                                      & Product                      & 78.87          & 92.72          & 97.72          & 99.42          \\
                                      &                              &                &                &                &                \\
Product                               & Clayton$(0.5)$               & 91.28          & 98.32          & 99.76          & 99.97          \\
                                      & Frank$(3)$                   & 62.98          & 99.03          & 99.93          & 100            \\
                                      & Gumbel-Hougaarad$(1.5)$                & 98.82          & 99.97          & 99.99          & 100            \\
                                      & Joe$(1.5)$                   & 66.39          & 85.64          & 95.2           & 98.62          \\
                                      & Normal$(0.4)$                & 100            & 100            & 100            & 100            \\
                                      & Product                      & \textbf{5.2}   & \textbf{4.99}  & \textbf{5.15}  & \textbf{5.97}  \\ \hline
\end{tabular}}
\end{table}
% \usepackage{multirow}
\begin{table}[H]
\centering
\caption{Percentage of rejection of $H_0$ for different bivariate copula models}\label{t4}
\scalebox{.9}{
\begin{tabular}{llllll}
\hline
\multirow{2}{*}{Copula   under $H_0$} & \multirow{2}{*}{True Copula} & \multicolumn{4}{l}{Sample Size}                                   \\ \cline{3-6} 
\multicolumn{1}{c}{}                                      & \multicolumn{1}{c}{}                             & 100            & 150            & 200            & 250            \\ \hline
Clayton$(2)$                                              & Clayton$(2)$                                     & \textbf{5.11}  & \textbf{4.93}  & \textbf{5.26}  & \textbf{4.99}  \\
                                                          & Frank$(5)$                                       & 94.81          & 99.69          & 99.98          & 100            \\
                                                          & Gumbel-Hougaarad$(2)$                                      & 95.81          & 99.71          & 100            & 100            \\
                                                          & Joe$(3)$                                         & 99.95          & 100            & 100            & 100            \\
                                                          & Normal$(0.7)$                                    & 70.81          & 90.19          & 97.62          & 99.19          \\
                                                          & Product                                          & 100            & 100            & 100            & 100            \\
                                                          &                                                  &                &                &                &                \\
Frank$(5)$                                                & Clayton$(2)$                                     & 97.47          & 99.75          & 100            & 100            \\
                                                          & Frank$(5)$                                       & \textbf{81.9}  & \textbf{93.07} & \textbf{97.92} & \textbf{99.37} \\
                                                          & Gumbel-Hougaarad$(2)$                                      & 7.36           & 10.96          & 13.2           & 16.56          \\
                                                          & Joe$(3)$                                         & 31.61          & 50.13          & 65.91          & 81.25          \\
                                                          & Normal$(0.7)$                                    & 10.56          & 21.08          & 32.08          & 41.89          \\
                                                          & Product                                          & 100            & 100            & 100            & 100            \\
                                                          &                                                  &                &                &                &                \\
Gumbel-Hougaarad$(2)$                                               & Clayton$(2)$                                     & 88.47          & 98.88          & 99.99          & 100            \\
                                                          & Frank$(5)$                                       & 7.41           & 10.77          & 14.74          & 17.88          \\
                                                          & Gumbel-Hougaarad$(2)$                                      & \textbf{5.13}  & \textbf{5.22}  & \textbf{4.83}  & \textbf{5.31}  \\
                                                          & Joe$(3)$                                         & 26.32          & 33.49          & 50.9           & 63.7           \\
                                                          & Normal$(0.7)$                                    & 5.08           & 8.9            & 13.72          & 19.2           \\
                                                          & Product                                          & 100            & 100            & 100            & 100            \\
                                                          &                                                  &                &                &                &                \\
Joe$(3)$                                                  & Clayton$(2)$                                     & 99.85          & 100            & 100            & 100            \\
                                                          & Frank$(5)$                                       & 19.76          & 33.25          & 49.75          & 60.54          \\
                                                          & Gumbel$(2)$                                      & 18.55          & 30.27          & 41.27          & 52.6           \\
                                                          & Joe$(3)$                                         & \textbf{4.77}           & \textbf{5.15 }          & \textbf{5.03}           & \textbf{5.11 }          \\
                                                          & Normal$(0.7)$                                    & 52.83 & 77.45 & 90.72 & 96.14 \\
                                                          & Product                                          & 100            & 100            & 100            & 100            \\
                                                          &                                                  &                &                &                &                \\
Normal$(0.7)$                                             & Clayton$(2)$                                     & 47.17          & 79.79          & 93.54          & 97.98          \\
                                                          & Frank$(5)$                                       & 26.08          & 37.19          & 48.03          & 56.66          \\
                                                          & Gumbel$(2)$                                      & 19.49          & 25.64          & 29.65          & 38.99          \\
                                                          & Joe$(3)$                                         & 68.4           & 89.71          & 96.63          & 99.05          \\
                                                          & Normal$(0.7)$                                    & \textbf{5.17}  & \textbf{5.19}  & \textbf{4.97}  & \textbf{5.08}  \\
                                                          & Product                                          & 100            & 100            & 100            & 100            \\
                                                          &                                                  &                &                &                &                \\
Product                                                   & Clayton$(2)$                                     & 100            & 100            & 100            & 100            \\
                                                          & Frank$(5)$                                       & 100            & 100            & 100            & 100            \\
                                                          & Gumbel-Hougaarad$(2)$                                      & 100            & 100            & 100            & 100            \\
                                                          & Joe$(3)$                                         & 100            & 100            & 100            & 100            \\
                                                          & Normal$(0.7)$                                    & 94.77          & 99.28          & 99.95          & 100            \\
                                                          & Product                                          & \textbf{4.97}  & \textbf{5.06}  & \textbf{4.84}  & \textbf{5.21}  \\ \hline
\end{tabular}}
\end{table}
% \usepackage{multirow}
\begin{table}[H]
\centering

\caption{Percentage of rejection of $H_0$ for different bivariate copula models}\label{t5}
\scalebox{.9}{\begin{tabular}{llllll}
\hline
\multirow{2}{*}{Copula   under $H_0$} & \multirow{2}{*}{True Copula} & \multicolumn{4}{l}{Sample Size}                                   \\ \cline{3-6} 
\multicolumn{1}{c}{}                                      & \multicolumn{1}{c}{}                             & 100            & 150            & 200           & 250            \\ \hline
Clayton$(6)$                                              & Clayton$(6)$                                     & \textbf{5.15}  & \textbf{5.06}  & \textbf{4.91} & \textbf{4.88}  \\
                                                          & Frank$(14)$                                      & 97.61          & 99.32          & 100           & 100            \\
                                                          & Gumbel-Hougaarad$(4)$                                      & 99.5           & 100            & 100           & 100            \\
                                                          & Joe$(7)$                                         & 100            & 100            & 100           & 100            \\
                                                          & Normal$(0.9)$                                    & 97.57          & 99.89          & 100           & 100            \\
                                                          & Product                                          & 100            & 100            & 100           & 100            \\
                                                          &                                                  &                &                &               &                \\
Frank$(14)$                                               & Clayton$(6)$                                     & 100            & 100            & 100           & 100            \\
                                                          & Frank$(14)$                                      & \textbf{5.02}  & \textbf{5.21}  & \textbf{4.93} & \textbf{4.87}  \\
                                                          & Gumbel-Hougaarad$(4)$                                      & 16.87          & 29.17          & 43.47         & 59.96          \\
                                                          & Joe$(7)$                                         & 59.16          & 78.61          & 88.58         & 95.55          \\
                                                          & Normal$(0.9)$                                    & 59.16          & 78.61          & 88.58         & 95.55          \\
                                                          & Product                                          & 100            & 100            & 100           & 100            \\
                                                          &                                                  &                &                &               &                \\
Gumbel-Hougaarad$(4)$                                               & Clayton$(6)$                                     & 86.12          & 97.8           & 99.62         & 99.99          \\
                                                          & Frank$(14)$                                      & 10.32          & 21.95          & 33.57         & 47.88          \\
                                                          & Gumbel-Hougaarad$(4)$                                      & \textbf{4.96}  & \textbf{5.22}  & \textbf{5.19} & \textbf{4.86}  \\
                                                          & Joe$(7)$                                         & 57.88          & 78.57          & 89.43         & 95.86          \\
                                                          & Normal$(0.9)$                                    & 6.77           & 8.67           & 11.28         & 17.07          \\
                                                          & Product                                          & 100            & 100            & 100           & 100            \\
                                                          &                                                  &                &                &               &                \\
Joe$(7)$                                                  & Clayton$(6)$                                     & 100            & 100            & 100           & 100            \\
                                                          & Frank$(14)$                                      & 20.86          & 46.27          & 67.1          & 82.65          \\
                                                          & Gumbel-Hougaarad$(4)$                                      & 32.6           & 65.06          & 83.02         & 92.35          \\
                                                          & Joe$(7)$                                         & \textbf{5.04}           & \textbf{5.02}          & \textbf{4.94}          & \textbf{4.91}           \\
                                                          & Normal$(0.9)$                                    & 69.29 & 93.75 & 99   & 99.82 \\
                                                          & Product                                          & 100            & 100            & 100           & 100            \\
                                                          &                                                  &                &                &               &                \\
Normal$(0.9)$                                             & Clayton$(6)$                                     & 21.01          & 87.3           & 99.82         & 99.98          \\
                                                          & Frank$(14)$                                      & 25.58          & 49.7           & 73.12         & 87.86          \\
                                                          & Gumbel-Hougaarad$(4)$                                      & 10.73          & 15.08          & 21.73         & 29.03          \\
                                                          & Joe$(7)$                                         & 82.06          & 97.5           & 99.7          & 99.99          \\
                                                          & Normal$(0.9)$                                    & \textbf{5.13}  & \textbf{4.85}  & \textbf{4.93} & \textbf{5.19}  \\
                                                          & Product                                          & 100            & 100            & 100           & 100            \\
                                                          &                                                  &                &                &               &                \\
Product                                                   & Clayton$(6)$                                     & 100            & 100            & 100           & 100            \\
                                                          & Frank$(14)$                                      & 100            & 100            & 100           & 100            \\
                                                          & Gumbel-Hougaarad$(4)$                                      & 100            & 100            & 100           & 100            \\
                                                          & Joe$(7)$                                         & 100            & 100            & 100           & 100            \\
                                                          & Normal$(0.9)$                                    & 100            & 100            & 100           & 100            \\
                                                          & Product                                          & \textbf{5.09}  & \textbf{5.04}  & \textbf{4.99} & \textbf{5.12}  \\ \hline
\end{tabular}}
\end{table}
% \usepackage{multirow}
\begin{table}[H]
\centering

\caption{Percentage of rejection of $H_0$ for different trivariate copula models}\label{t6}
\scalebox{.9}{\begin{tabular}{llllll}
\hline
\multirow{2}{*}{Copula   under $H_0$} & \multirow{2}{*}{True Copula} & \multicolumn{4}{l}{Sample Size}                                   \\ \cline{3-6} 
\multicolumn{1}{c}{}                                      & \multicolumn{1}{c}{}                             & 100            & 150            & 200            & 250            \\ \hline
Clayton$(0.5)$                                            & Clayton$(0.5)$                                   & \textbf{4.78}  & \textbf{4.83}  & \textbf{5.1}   & \textbf{4.89}  \\
                                                          & Frank$(3)$                                       & 41             & 70.7           & 88.65          & 96.03          \\
                                                          & Gumbel-Hougaarad$(1.5)$                                    & 50.46          & 78.26          & 92.66          & 97.55          \\
                                                          & Joe$(1.5)$                                       & 83.41          & 96.77          & 99.45          & 99.94          \\
                                                          & Normal$(0.1,0.2,0.3)$                            & 62.8           & 80.15          & 91.54          & 96.94          \\
                                                          & Product                                          & 99.79          & 99.99          & 100            & 100            \\
                                                          &                                                  &                &                &                &                \\
Frank$(3)$                                                & Clayton$(0.5)$                                   & 63.19          & 86.03          & 94.62          & 98.14          \\
                                                          & Frank$(3)$                                       & \textbf{4.94}  & \textbf{4.91}  & \textbf{5.09}  & \textbf{5.05}  \\
                                                          & Gumbel-Hougaarad$(1.5)$                                    & 8.73           & 9.85           & 10.46          & 13.83          \\
                                                          & Joe$(1.5)$                                       & 81.9           & 93.07          & 97.92          & 99.37          \\
                                                          & Normal$(0.1,0.2,0.3)$                            & 89.47          & 97.95          & 99.6           & 99.93          \\
                                                          & Product                                          & 100            & 100            & 100            & 100            \\
                                                          &                                                  &                &                &                &                \\
Gumbel-Hougaarad$(1.5)$                                             & Clayton$(0.5)$                                   & 57.05          & 82.82          & 94.97          & 98.33          \\
                                                          & Frank$(3)$                                       & 4.85           & 4.96           & 5.48           & 6.83           \\
                                                          & Gumbel-Hougaarad$(1.5)$                                    & \textbf{5.14}  & \textbf{5.21}  & \textbf{4.89}  & \textbf{4.96}  \\
                                                          & Joe$(1.5)$                                       & 74.33          & 89.29          & 95.52          & 98.31          \\
                                                          & Normal$(0.1,0.2,0.3)$                            & 88.47          & 97.21          & 99.55          & 99.93          \\
                                                          & Product                                          & 99.4           & 99.99          & 100            & 100            \\
                                                          &                                                  &                &                &                &                \\
Joe$(1.5)$                                                & Clayton$(0.5)$                                   & 79.73          & 93.73          & 98.63          & 99.77          \\
                                                          & Frank$(3)$                                       & 68.91          & 88.06          & 95.55          & 98.69          \\
                                                          & Gumbel-Hougaarad$(1.5)$                                    & 66.71          & 84.56          & 92.61          & 97.17          \\
                                                          & Joe$(1.5)$                                       & \textbf{4.99}           & \textbf{4.83}           & \textbf{5.18}           & \textbf{4.91}           \\
                                                          & Normal$(0.1,0.2,0.3)$                            & 17.47 & 23.55 & 36.72 & 49.04 \\
                                                          & Product                                          & 93.09          & 99.17          & 99.92          & 99.98          \\
                                                          &                                                  &                &                &                &                \\
Normal$(0.1,0.2,0.3)$                                     & Clayton$(0.5)$                                   & 59.44          & 80.12          & 89.17          & 95.21          \\
                                                          & Frank$(3)$                                       & 83.75          & 96.26          & 99.3           & 99.85          \\
                                                          & Gumbel-Hougaarad$(1.5)$                                    & 82.16          & 96.06          & 99.18          & 99.8           \\
                                                          & Joe$(1.5)$                                       & 16.99          & 27.04          & 39.63          & 55.51          \\
                                                          & Normal$(0.1,0.2,0.3)$                            & \textbf{5.19}  & \textbf{4.95}  & \textbf{5.12}  & \textbf{5.08}  \\
                                                          & Product                                          & 80.13          & 93.05          & 97.41          & 99.47          \\
                                                          &                                                  &                &                &                &                \\
Product                                                   & Clayton$(0.5)$                                   & 99.45          & 99.97          & 100            & 100            \\
                                                          & Frank$(3)$                                       & 80.61          & 100            & 100            & 100            \\
                                                          & Gumbel-Hougaarad$(1.5)$                                    & 100            & 100            & 100            & 100            \\
                                                          & Joe$(1.5)$                                       & 83.41          & 96.77          & 99.45          & 99.97          \\
                                                          & Normal$(0.1,0.2,0.3)$                            & 75.22          & 91.04          & 96.68          & 98.86          \\
                                                          & Product                                          & \textbf{4.98}  & \textbf{5.09}  & \textbf{4.99}  & \textbf{4.93}  \\ \hline
\end{tabular}}
\end{table}
% \usepackage{multirow}
\begin{table}[H]
\centering

\caption{Percentage of rejection of $H_0$ for different trivariate copula models}\label{t7}
\scalebox{.9}{\begin{tabular}{llllll}
\hline
\multirow{2}{*}{Copula   under $H_0$} & \multirow{2}{*}{True Copula} & \multicolumn{4}{l}{Sample Size}                                   \\ \cline{3-6} 
\multicolumn{1}{c}{}                                      & \multicolumn{1}{c}{}                             & 100           & 150           & 200           & 250           \\ \hline
Clayton$(2)$                                              & Clayton$(2)$                                     & \textbf{5.03} & \textbf{4.98} & \textbf{4.91} & \textbf{4.99} \\
                                                          & Frank$(5)$                                       & 94.81         & 99.69         & 99.98         & 100           \\
                                                          & Gumbel-Hougaarad$(2)$                                      & 99.38         & 99.99         & 100           & 100           \\
                                                          & Joe$(3)$                                         & 100           & 100           & 100           & 100           \\
                                                          & Normal$(0.4,0.5,0.6)$                            & 99.96         & 100           & 100           & 100           \\
                                                          & Product                                          & 100           & 100           & 100           & 100           \\
                                                          &                                                  &               &               &               &               \\
Frank$(5)$                                                & Clayton$(2)$                                     & 99.88         & 100           & 100           & 100           \\
                                                          & Frank$(5)$                                       & \textbf{5.11} & \textbf{5.17} & \textbf{4.84} & \textbf{5.15} \\
                                                          & Gumbel-Hougaarad$(2)$                                      & 9.52          & 14.29         & 21.34         & 30.71         \\
                                                          & Joe$(3)$                                         & 49.63         & 77.6          & 90.36         & 97.24         \\
                                                          & Normal$(0.4,0.5,0.6)$                            & 62.34         & 79.4          & 90.48         & 96.36         \\
                                                          & Product                                          & 100           & 100           & 100           & 100           \\
                                                          &                                                  &               &               &               &               \\
Gumbel-Hougaarad$(2)$                                               & Clayton$(2)$                                     & 97.33         & 99.86         & 100           & 100           \\
                                                          & Frank$(5)$                                       & 9.25          & 11.76         & 18.89         & 25.86         \\
                                                          & Gumbel-Hougaarad$(2)$                                      & \textbf{4.97} & \textbf{4.87} & \textbf{4.91} & \textbf{4.93} \\
                                                          & Joe$(3)$                                         & 39.86         & 58.63         & 72.5          & 85.18         \\
                                                          & Normal$(0.4,0.5,0.6)$                            & 63.75         & 83.55         & 94.43         & 98.73         \\
                                                          & Product                                          & 100           & 100           & 100           & 100           \\
                                                          &                                                  &               &               &               &               \\
Joe$(3)$                                                  & Clayton$(2)$                                     & 100           & 100           & 100           & 100           \\
                                                          & Frank$(5)$                                       & 29.85         & 51.43         & 70.19         & 83.57         \\
                                                          & Gumbel-Hougaarad$(2)$                                      & 28.93         & 49.53         & 64.89         & 77.01         \\
                                                          & Joe$(3)$                                         & \textbf{5.01} & \textbf{5.18} & \textbf{4.96} & \textbf{5.11} \\
                                                          & Normal$(0.4,0.5,0.6)$                            & 81.11         & 97.16         & 99.7          & 99.98         \\
                                                          & Product                                          & 100           & 100           & 100           & 100           \\
                                                          &                                                  &               &               &               &               \\
Normal$(0.4,0.5,0.6)$                                     & Clayton$(2)$                                     & 99.61         & 100           & 100           & 100           \\
                                                          & Frank$(5)$                                       & 43.56         & 72.26         & 89.59         & 94.96         \\
                                                          & Gumbel-Hougaarad$(2)$                                      & 53.71         & 81.1          & 93.52         & 97.99         \\
                                                          & Joe$(3)$                                         & 87.08         & 98.83         & 99.93         & 99.99         \\
                                                          & Normal$(0.4,0.5,0.6)$                            & \textbf{5.02} & \textbf{5.19} & \textbf{4.81} & \textbf{5.29} \\
                                                          & Product                                          & 100           & 100           & 100           & 100           \\
                                                          &                                                  &               &               &               &               \\
Product                                                   & Clayton$(2)$                                     & 100           & 100           & 100           & 100           \\
                                                          & Frank$(5)$                                       & 100           & 100           & 100           & 100           \\
                                                          & Gumbel-Hougaarad$(2)$                                      & 100           & 100           & 100           & 100           \\
                                                          & Joe$(3)$                                         & 100           & 100           & 100           & 100           \\
                                                          & Normal$(0.4,0.5,0.6)$                            & 100           & 100           & 100           & 100           \\
                                                          & Product                                          & \textbf{4.97} & \textbf{5.11} & \textbf{4.85} & \textbf{5.04} \\ \hline
\end{tabular}}
\end{table}
\begin{table}[H]
\centering

\caption{Percentage of rejection of $H_0$ for different trivariate copula models}\label{t8}
\scalebox{.9}{\begin{tabular}{llllll}
\hline
\multirow{2}{*}{Copula   under $H_0$} & \multirow{2}{*}{True Copula} & \multicolumn{4}{l}{Sample Size}                                   \\ \cline{3-6} 
\multicolumn{1}{c}{}                                      & \multicolumn{1}{c}{}                             & 100           & 150           & 200           & 250           \\ \hline
Clayton$(6)$                                              & Clayton$(6)$                                     & \textbf{4.99} & \textbf{4.93} & \textbf{4.89} & \textbf{5.01} \\
                                                          & Frank$(14)$                                      & 99.71         & 100           & 100           & 100           \\
                                                          & Gumbel$(4)$                                      & 99.98         & 100           & 100           & 100           \\
                                                          & Joe$(7)$                                         & 100           & 100           & 100           & 100           \\
                                                          & Normal$(0.7,0.8,0.9)$                            & 100           & 100           & 100           & 100           \\
                                                          & Product                                          & 100           & 100           & 100           & 100           \\
                                                          &                                                  &               &               &               &               \\
Frank$(14)$                                               & Clayton$(6)$                                     & 100           & 100           & 100           & 100           \\
                                                          & Frank$(14)$                                      & \textbf{5.23} & \textbf{4.97} & \textbf{4.86} & \textbf{5.09} \\
                                                          & Gumbel-Hougaarad$(4)$                                      & 21.68         & 40.24         & 61.75         & 78.87         \\
                                                          & Joe$(7)$                                         & 90.55         & 97.01         & 99.51         & 99.98         \\
                                                          & Normal$(0.7,0.8,0.9)$                            & 97.87         & 99.82         & 100           & 100           \\
                                                          & Product                                          & 100           & 100           & 100           & 100           \\
                                                          &                                                  &               &               &               &               \\
Gumbel-Hougaarad$(4)$                                               & Clayton$(6)$                                     & 84.46         & 99.9          & 100           & 100           \\
                                                          & Frank$(14)$                                      & 14.19         & 31.04         & 50.42         & 67.88         \\
                                                          & Gumbel-Hougaarad$(4)$                                      & \textbf{5.34} & \textbf{4.95} & \textbf{4.9}  & \textbf{5.02} \\
                                                          & Joe$(7)$                                         & 73.51         & 92.36         & 97.86         & 99.6          \\
                                                          & Normal$(0.7,0.8,0.9)$                            & 86.57         & 98.19         & 99.85         & 100           \\
                                                          & Product                                          & 100           & 100           & 100           & 100           \\
                                                          &                                                  &               &               &               &               \\
Joe$(7)$                                                  & Clayton$(6)$                                     & 100           & 100           & 100           & 100           \\
                                                          & Frank$(14)$                                      & 31.14         & 64.39         & 84.83         & 94.35         \\
                                                          & Gumbel-Hougaarad$(4)$                                      & 54.3          & 86.37         & 95.85         & 98.94         \\
                                                          & Joe$(7)$                                         & \textbf{4.95} & \textbf{5.28} & \textbf{5.05} & \textbf{4.98} \\
                                                          & Normal$(0.7,0.8,0.9)$                            & 88.43         & 99.16         & 99.93         & 99.99         \\
                                                          & Product                                          & 100           & 100           & 100           & 100           \\
                                                          &                                                  &               &               &               &               \\
Normal$(0.7,0.8,0.9)$                                     & Clayton$(6)$                                     & 99.97         & 100           & 100           & 100           \\
                                                          & Frank$(14)$                                      & 94.31         & 99.46         & 100           & 100           \\
                                                          & Gumbel-Hougaarad$(4)$                                      & 60.5          & 95.81         & 99.77         & 100           \\
                                                          & Joe$(7)$                                         & 98.1          & 99.98         & 100           & 100           \\
                                                          & Normal$(0.7,0.8,0.9)$                            & \textbf{4.76} & \textbf{5.32} & \textbf{5.24} & \textbf{4.89} \\
                                                          & Product                                          & 100           & 100           & 100           & 100           \\
                                                          &                                                  &               &               &               &               \\
Product                                                   & Clayton$(6)$                                     & 100           & 100           & 100           & 100           \\
                                                          & Frank$(14)$                                      & 100           & 100           & 100           & 100           \\
                                                          & Gumbel-Hougaarad$(4)$                                      & 100           & 100           & 100           & 100           \\
                                                          & Joe$(7)$                                         & 100           & 100           & 100           & 100           \\
                                                          & Normal$(0.7,0.8,0.9)$                            & 100           & 100           & 100           & 100           \\
                                                          & Product                                          & \textbf{5.04} & \textbf{4.89} & \textbf{4.99} & \textbf{5.08} \\ \hline
\end{tabular}}
\end{table}
\bibliographystyle{IEEEtran}
\bibliography{references}

\begin{thebibliography}{10}
\providecommand{\url}[1]{#1}
\csname url@samestyle\endcsname
\providecommand{\newblock}{\relax}
\providecommand{\bibinfo}[2]{#2}
\providecommand{\BIBentrySTDinterwordspacing}{\spaceskip=0pt\relax}
\providecommand{\BIBentryALTinterwordstretchfactor}{4}
\providecommand{\BIBentryALTinterwordspacing}{\spaceskip=\fontdimen2\font plus
\BIBentryALTinterwordstretchfactor\fontdimen3\font minus
  \fontdimen4\font\relax}
\providecommand{\BIBforeignlanguage}[2]{{%
\expandafter\ifx\csname l@#1\endcsname\relax
\typeout{** WARNING: IEEEtran.bst: No hyphenation pattern has been}%
\typeout{** loaded for the language `#1'. Using the pattern for}%
\typeout{** the default language instead.}%
\else
\language=\csname l@#1\endcsname
\fi
#2}}
\providecommand{\BIBdecl}{\relax}
\BIBdecl

\bibitem{a1}
P.~Perrone, ``Markov categories and entropy,'' \emph{IEEE Transactions on
  Information Theory}, vol.~70, no.~3, 2024.

\bibitem{a2}
X.~Wang, S.~Zhang, and T.~Li, ``A quantum algorithm framework for discrete
  probability distributions with applications to r{\'e}nyi entropy
  estimation,'' \emph{IEEE Transactions on Information Theory}, vol.~70, no.~5,
  2024.

\bibitem{a3}
A.~Ali, S.~Anam, and M.~M. Ahmed, ``Shannon entropy in artificial intelligence
  and its applications based on information theory,'' \emph{Journal of Applied
  and Emerging Sciences}, vol.~13, no.~1, pp. 09--17, 2023.

\bibitem{a4}
L.~Yan and X.~Ge, ``Entropy-based energy dissipation analysis of mobile
  communication systems,'' \emph{IEEE Transactions on Mobile Computing},
  vol.~23, no.~6, 2024 (to appear).

\bibitem{a5}
R.~Ren, Y.~Li, Q.~Sun, X.~Xie, L.~Liu, and D.~W. Gao, ``Digital twin assisted
  economic dispatch for energy internet with information entropy,'' \emph{IEEE
  Transactions on Automation Science and Engineering}, 2024.

\bibitem{hosseini2021discussion}
T.~Hosseini and M.~J. Nooghabi, ``Discussion about inaccuracy measure in
  information theory using co-copula and copula dual functions,'' \emph{Journal
  of Multivariate Analysis}, vol. 183, p. 104725, 2021.

\bibitem{f5}
F.~Foroghi, S.~Tahmasebi, M.~Afshari, and F.~Lak, ``Extensions of fractional
  cumulative residual entropy with applications,'' \emph{Communications in
  Statistics-Theory and Methods}, vol.~52, no.~20, pp. 7350--7369, 2023.

\bibitem{f6}
S.~Saha and S.~Kayal, ``Extended fractional cumulative past and paired
  $\phi$-entropy measures,'' \emph{Physica A: Statistical Mechanics and its
  Applications}, vol. 614, p. 128552, 2023.

\bibitem{shannon1948mathematical}
C.~E. Shannon, ``A mathematical theory of communication,'' \emph{The Bell
  system technical journal}, vol.~27, no.~3, pp. 379--423, 1948.

\bibitem{rao2004cumulative}
M.~Rao, Y.~Chen, B.~C. Vemuri, and F.~Wang, ``Cumulative residual entropy: a
  new measure of information,'' \emph{IEEE transactions on Information Theory},
  vol.~50, no.~6, pp. 1220--1228, 2004.

\bibitem{di2009cumulative}
A.~Di~Crescenzo and M.~Longobardi, ``On cumulative entropies,'' \emph{Journal
  of statistical planning and inference}, vol. 139, no.~12, pp. 4072--4087,
  2009.

\bibitem{renyi1961measures}
A.~R{\'e}nyi, ``On measures of entropy and information,'' in \emph{Proceedings
  of the Fourth Berkeley Symposium on Mathematical Statistics and Probability,
  Volume 1: Contributions to the Theory of Statistics}.\hskip 1em plus 0.5em
  minus 0.4em\relax The Regents of the university of California Press, 1961.

\bibitem{tsallis1988possible}
C.~Tsallis, ``Possible generalization of boltzmann-gibbs statistics,''
  \emph{Journal of Statistical Physics}, vol.~52, pp. 479--487, 1988.

\bibitem{di2007weighted}
A.~Di~Crescenzo and M.~Longobardi, ``On weighted residual and past entropies,''
  \emph{arXiv preprint math/0703489}, 2007.

\bibitem{mathai2007pathway}
A.~Mathai and H.~J. Haubold, ``Pathway model, superstatistics, tsallis
  statistics, and a generalized measure of entropy,'' \emph{Physica A:
  Statistical Mechanics and its Applications}, vol. 375, no.~1, pp. 110--122,
  2007.

\bibitem{golomb1966information}
S.~Golomb, ``The information generating function of a probability
  distribution,'' \emph{IEEE Transactions on Information Theory}, vol.~12,
  no.~1, pp. 75--77, 1966.

\bibitem{guiasu1985relative}
S.~Guiasu and C.~Reischer, ``The relative information generating function,''
  \emph{Information Sciences}, vol.~35, no.~3, pp. 235--241, 1985.

\bibitem{kullback1951information}
S.~Kullback and R.~A. Leibler, ``On information and sufficiency,'' \emph{The
  Annals of Mathematical Statistics}, vol.~22, no.~1, pp. 79--86, 1951.

\bibitem{papaioannou2007some}
T.~Papaioannou, K.~Ferentinos, and C.~Tsairidis, ``Some information theoretic
  ideas useful in statistical inference,'' \emph{Methodology and Computing in
  Applied Probability}, vol.~9, pp. 307--323, 2007.

\bibitem{smitha2019dynamic}
S.~Smitha, G.~Rajesh, and A.~Baby, ``On dynamic cumulative past entropy
  generating function,'' \emph{Think India Journal}, vol.~22, no.~14, pp.
  9400--9406, 2019.

\bibitem{smitha2023dynamic}
S.~Smitha, S.~K. Kattumannil, and E.~Sreedevi, ``Dynamic cumulative residual
  entropy generating function and its properties,'' \emph{Communications in
  Statistics-Theory and Methods}, vol.~53, no.~16, pp. 5890--5909, 2024.

\bibitem{saha2024general}
S.~Saha and S.~Kayal, ``General weighted information and relative information
  generating functions with properties,'' \emph{IEEE Transactions on
  Information Theory}, vol.~70, no.~8, 2024.

\bibitem{ubriaco2009entropies}
M.~R. Ubriaco, ``Entropies based on fractional calculus,'' \emph{Physics
  Letters A}, vol. 373, no.~30, pp. 2516--2519, 2009.

\bibitem{xiong2019fractional}
H.~Xiong, P.~Shang, and Y.~Zhang, ``Fractional cumulative residual entropy,''
  \emph{Communications in Nonlinear Science and Numerical Simulation}, vol.~78,
  p. 104879, 2019.

\bibitem{kayid2022some}
M.~Kayid and M.~Shrahili, ``Some further results on the fractional cumulative
  entropy,'' \emph{Entropy}, vol.~24, no.~8, p. 1037, 2022.

\bibitem{f1}
G.~Jumarie, ``Derivation of an amplitude of information in the setting of a new
  family of fractional entropies,'' \emph{Information Sciences}, vol. 216, pp.
  113--137, 2012.

\bibitem{f2}
A.~Karci, ``Fractional order entropy: New perspectives,'' \emph{Optik}, vol.
  127, no.~20, pp. 9172--9177, 2016.

\bibitem{f3}
A.~M. Lopes and J.~A.~T. Machado, ``A review of fractional order entropies,''
  \emph{Entropy}, vol.~22, no.~12, p. 1374, 2020.

\bibitem{f4}
A.~Di~Crescenzo, S.~Kayal, and A.~Meoli, ``Fractional generalized cumulative
  entropy and its dynamic version,'' \emph{Communications in Nonlinear Science
  and Numerical Simulation}, vol. 102, p. 105899, 2021.

\bibitem{nadarajah2005expressions}
S.~Nadarajah and K.~Zografos, ``Expressions for r{\'e}nyi and shannon entropies
  for bivariate distributions,'' \emph{Information Sciences}, vol. 170, no.~2,
  pp. 173--189, 2005.

\bibitem{ebrahimi2007multivariate}
N.~Ebrahimi, S.~Kirmani, and E.~S. Soofi, ``Multivariate dynamic information,''
  \emph{Journal of Multivariate Analysis}, vol.~98, no.~2, pp. 328--349, 2007.

\bibitem{rajesh2009bivariate}
G.~Rajesh, A.~Sathar, and K.~M. Nair, ``Bivariate extension of residual entropy
  and some characterization results,'' \emph{Journal of Indian Statistical
  Association}, vol.~47, no.~1, pp. 91--107, 2009.

\bibitem{rajesh2014bivariate}
G.~Rajesh, E.~Abdul-Sathar, K.~M. Nair, and K.~Reshmi, ``Bivariate extension of
  dynamic cumulative residual entropy,'' \emph{Statistical Methodology},
  vol.~16, pp. 72--82, 2014.

\bibitem{asadi2007dynamic}
M.~Asadi and Y.~Zohrevand, ``On the dynamic cumulative residual entropy,''
  \emph{Journal of Statistical Planning and Inference}, vol. 137, no.~6, pp.
  1931--1941, 2007.

\bibitem{c2017bivariate}
A.~Kundu and C.~Kundu, ``Bivariate extension of (dynamic) cumulative past
  entropy,'' \emph{Communications in Statistics-Theory and Methods}, vol.~46,
  no.~9, pp. 4163--4180, 2017.

\bibitem{sklar1959fonctions}
M.~Sklar, ``Fonctions de repartition an dimensions et leurs marges,''
  \emph{Publ. inst. statist. univ. Paris}, vol.~8, pp. 229--231, 1959.

\bibitem{nelsen2007introduction}
R.~B. Nelsen, \emph{An introduction to copulas}.\hskip 1em plus 0.5em minus
  0.4em\relax Springer Science \& Business Media, 2006.

\bibitem{durante2016principles}
F.~Durante and C.~Sempi, \emph{Principles of copula theory}.\hskip 1em plus
  0.5em minus 0.4em\relax CRC Press, 2016.

\bibitem{chesneau2022note}
C.~Chesneau, ``A note on a simple polynomial-sine copula,'' \emph{Asian Journal
  of Mathematics and Applications}, vol.~2, pp. 1--14, 2022.

\bibitem{zachariah2024new}
S.~G. Zachariah, M.~Arshad, and A.~K. Pathak, ``A new class of copulas having
  dependence range larger than fgm-type copulas,'' \emph{Statistics \&
  Probability Letters}, vol. 206, p. 109988, 2024.

\bibitem{ma2011mutual}
J.~Ma and Z.~Sun, ``Mutual information is copula entropy,'' \emph{Tsinghua
  Science \& Technology}, vol.~16, no.~1, pp. 51--54, 2011.

\bibitem{sunoj2023survival}
S.~Sunoj and N.~U. Nair, ``Survival copula entropy and dependence in bivariate
  distributions: Accepted-february 2023,'' \emph{REVSTAT-Statistical Journal},
  2023.

\bibitem{ce1}
N.~Zhao and W.~T. Lin, ``A copula entropy approach to correlation measurement
  at the country level,'' \emph{Applied Mathematics and Computation}, vol. 218,
  no.~2, pp. 628--642, 2011.

\bibitem{ce2}
Z.~Hao and V.~P. Singh, ``Integrating entropy and copula theories for
  hydrologic modeling and analysis,'' \emph{Entropy}, vol.~17, no.~4, pp.
  2253--2280, 2015.

\bibitem{ce3}
V.~P. Singh and L.~Zhang, ``Copula--entropy theory for multivariate stochastic
  modeling in water engineering,'' \emph{Geoscience Letters}, vol.~5, no.~1,
  pp. 1--17, 2018.

\bibitem{saha2023copula}
S.~Saha and S.~Kayal, ``Copula-based extropy measures, properties and
  dependence in bivariate distributions,'' \emph{arXiv preprint
  arXiv:2311.08061}, 2023.

\bibitem{cuadras2009constructing}
C.~M. Cuadras, ``Constructing copula functions with weighted geometric means,''
  \emph{Journal of Statistical Planning and Inference}, vol. 139, no.~11, pp.
  3766--3772, 2009.

\bibitem{schmid2010copula}
F.~Schmid, R.~Schmidt, T.~Blumentritt, S.~Gai{\ss}er, and M.~Ruppert,
  ``Copula-based measures of multivariate association,'' in \emph{Copula Theory
  and Its Applications}, P.~Jaworski, F.~Durante, W.~K. H{\"a}rdle, and
  T.~Rychlik, Eds.\hskip 1em plus 0.5em minus 0.4em\relax Berlin, Heidelberg:
  Springer Berlin Heidelberg, 2010, pp. 209--236.

\bibitem{bedHo2016multivariate}
J.~Bedő and C.~S. Ong, ``Multivariate spearman's {$\rho$} for aggregating
  ranks using copulas,'' \emph{Journal of Machine Learning Research}, vol.~17,
  no. 201, pp. 1--30, 2016.

\bibitem{shaked2007stochastic}
M.~Shaked and J.~G. Shanthikumar, \emph{Stochastic orders}.\hskip 1em plus
  0.5em minus 0.4em\relax Springer, 2007.

\bibitem{bartoszewicz1997dispersive}
J.~Bartoszewicz, ``Dispersive functions and stochastic orders,''
  \emph{Applicationes Mathematicae}, vol.~24, no.~4, pp. 429--444, 1997.

\bibitem{zhang2013some}
K.~Zhang, J.~Lin, and C.~Huang, ``Some new results on weighted geometric mean
  for copulas,'' \emph{International Journal of Uncertainty, Fuzziness and
  Knowledge-Based Systems}, vol.~21, no.~02, pp. 277--288, 2013.

\bibitem{kufner1977function}
A.~Kufner, O.~John, and S.~Fucik, ``Function spaces, noordhoff int,''
  \emph{Pub. Leyden, Academia Prague}, 1977.

\bibitem{finner1992generalization}
H.~Finner, ``A generalization of holder's inequality and some probability
  inequalities,'' \emph{The Annals of Probability}, vol.~20, no.~4, pp.
  1893--1901, 1992.

\bibitem{segers2017empirical}
J.~Segers, M.~Sibuya, and H.~Tsukahara, ``The empirical beta copula,''
  \emph{Journal of Multivariate Analysis}, vol. 155, pp. 35--51, 2017.

\bibitem{sancetta2004bernstein}
A.~Sancetta and S.~Satchell, ``The bernstein copula and its applications to
  modeling and approximations of multivariate distributions,''
  \emph{Econometric theory}, vol.~20, no.~3, pp. 535--562, 2004.

\bibitem{ec1}
J.~Kiefer, ``On large deviations of the empiric d.f. of vector chance variables
  and a law of the iterated logarithm.'' \emph{Pacific J. Math.}, vol.~11,
  no.~2, pp. 649--660, 1961.

\bibitem{ec2}
G.~R. Shorack and J.~A. Wellner, \emph{Empirical processes with applications to
  statistics}.\hskip 1em plus 0.5em minus 0.4em\relax SIAM, 2009.

\bibitem{ec3}
P.~Janssen, J.~Swanepoel, and N.~Veraverbeke, ``Large sample behavior of the
  bernstein copula estimator,'' \emph{Journal of Statistical Planning and
  Inference}, vol. 142, no.~5, pp. 1189--1197, 2012.

\bibitem{ec4}
J.~M. Gonz{\'a}lez-Barrios and R.~Hoyos-Arg{\"u}elles, ``Estimating
  checkerboard approximations with sample d-copulas,'' \emph{Communications in
  Statistics-Simulation and Computation}, vol.~50, no.~12, pp. 3992--4027,
  2021.

\bibitem{baratpour2012testing}
S.~Baratpour and A.~H. Rad, ``Testing goodness-of-fit for exponential
  distribution based on cumulative residual entropy,'' \emph{Communications in
  Statistics-Theory and Methods}, vol.~41, no.~8, pp. 1387--1396, 2012.

\bibitem{yela2018estimating}
J.~P. Yela and J.~R.~T. Cuevas, ``Estimating the gumbel-barnett copula
  parameter of dependence,'' \emph{Revista Colombiana de Estad{\'\i}stica},
  vol.~41, no.~1, pp. 53--73, 2018.

\bibitem{GF2}
V.~Panchenko, ``Goodness-of-fit test for copulas,'' \emph{Physica A:
  Statistical Mechanics and its Applications}, vol. 355, no.~1, pp. 176--182,
  2005.

\bibitem{gF1}
C.~Genest, B.~R{\'e}millard, and D.~Beaudoin, ``Goodness-of-fit tests for
  copulas: A review and a power study,'' \emph{Insurance: Mathematics and
  economics}, vol.~44, no.~2, pp. 199--213, 2009.

\bibitem{gf3}
I.~Kojadinovic, J.~Yan, and M.~Holmes, ``Fast large-sample goodness-of-fit
  tests for copulas,'' \emph{Statistica Sinica}, pp. 841--871, 2011.

\end{thebibliography}



\end{document}